\documentclass[twocolumn]{aastex63}

\usepackage{graphicx}
\usepackage{placeins}
\usepackage{amsmath}

\def\etal{{et~al.\null}}
\newcommand{\OII}{[\ion{O}{2}]}
\newcommand{\OIII}{[\ion{O}{3}]}

\newcommand{\NII}{[\ion{N}{2}]}
\newcommand{\SII}{[\ion{S}{2}]}

\shorttitle{HETDEX [O~III] Emitters}
\shortauthors{Indahl et al.}
\begin{document}

\title{HETDEX [OIII] Emitters I:  A spectroscopically selected low-redshift population of low-mass, low-metallicity galaxies}

\correspondingauthor{Briana Indahl}
\email{bindahl@utexas.edu}

\author{Briana Indahl}
\affiliation{Astronomy Department, University of Texas at Austin, Austin, TX 78712}
\nocollaboration{1}

\author{Greg Zeimann}
\affiliation{Hobby Eberly Telescope, University of Texas, Austin, Austin, TX, 78712}
\nocollaboration{1}

\author{Gary J. Hill}
\affiliation{Astronomy Department, University of Texas at Austin, Austin, TX 78712}
\affiliation{McDonald Observatory, University of Texas, Austin, TX, 78712}
\nocollaboration{1}

\author{William P. Bowman}
\affiliation{Department of Astronomy \& Astrophysics, Pennsylvania State University, University Park, PA 16802}
\affiliation{Institute for Gravitation \& the Cosmos, Pennsylvania State University, University Park, PA, 16802}
\nocollaboration{1}

\author{Robin Ciardullo}
\affiliation{Department of Astronomy \& Astrophysics, Pennsylvania State University, University Park, PA 16802}
\affiliation{Institute for Gravitation \& the Cosmos, Pennsylvania State University, University Park, PA, 16802}
\nocollaboration{1}

\author{Niv Drory}
%\affiliation{Astronomy Department, University of Texas at Austin, TX 78712}
\affiliation{McDonald Observatory, University of Texas, Austin, TX, 78712}
\nocollaboration{1}

\author{Eric Gawiser}
\affiliation{Department of Physics and Astronomy, Rutgers, The State University of New Jersey, 136 Frelinghuysen Road, Piscataway, NJ 08854, USA}
\nocollaboration{1}

\author{Ulrich Hopp}
\affiliation{University Observatory, Fakult\"at f\"ur Physik, Ludwig-Maximilians-Universit\"at, Scheinerstr. 1, 81679 Munich, Germany}
\affiliation{Max-Planck-Institut f{\"u}r extraterrestrische Physik, 
Giessenbachstrasse 1, 85748 Garching, Germany}
\nocollaboration{1}

\author{Steven Janowiecki}
\affiliation{McDonald Observatory, University of Texas, Austin, TX, 78712}
\nocollaboration{1}

\author{Michael Boylan-Kolchin}
\affiliation{Astronomy Department, University of Texas at Austin, Austin, TX 78712}
\nocollaboration{1}

\author{Erin Mentuch Cooper}
\affiliation{Astronomy Department, University of Texas at Austin, Austin, TX 78712}
\nocollaboration{1}

\author{Dustin Davis}
\affiliation{Astronomy Department, University of Texas at Austin, Austin, TX 78712}
\nocollaboration{1}

\author{Daniel Farrow}
\affiliation{Max-Planck-Institut f{\"u}r extraterrestrische Physik, 
Giessenbachstrasse 1, 85748 Garching, Germany}
\nocollaboration{1}

\author{Steven Finkelstein}
\affiliation{Astronomy Department, University of Texas at Austin, Austin, TX 78712}
\nocollaboration{1}

\author{Caryl Gronwall}
\affiliation{Department of Astronomy \& Astrophysics, Pennsylvania State University, University Park, PA 16802}
\affiliation{Institute for Gravitation \& the Cosmos, Pennsylvania State University, University Park, PA, 16802}
\nocollaboration{1}

\author{Andreas Kelz}
\affiliation{Leibniz-Institut fuer Astrophysik Potsdam (AIP), An der Sternwarte 16 D-14482 Potsdam, Germany}
\nocollaboration{1}

\author{Kristen B. W. McQuinn}
\affiliation{Rutgers University, Department of Physics and Astronomy, 136 Frelinghuysen Road, Piscataway, NJ 08854, USA}
\nocollaboration{1}

\author{Don Schneider}
\affiliation{Department of Astronomy \& Astrophysics, Pennsylvania State University, University Park, PA 16802}
\affiliation{Institute for Gravitation \& the Cosmos, Pennsylvania State University, University Park, PA, 16802}
\nocollaboration{1}

\author{Sarah E. Tuttle}
\affiliation{University of Washington, Seattle, 3910 15th Ave. NE, Room C319, Seattle, WA 98195-0002, USA}
\nocollaboration{1}



\begin{abstract}

We assemble a sample of 17 low metallicity (7.45 $<$ log(O/H)+12 $<$ 8.12) galaxies with $z \lesssim 0.1$ found spectroscopically, without photometric pre-selection, in early data from the Hobby Eberly Telescope Dark Energy Experiment (HETDEX)\null.  Star forming galaxies that occupy the lowest mass and metallicity end of the mass-metallicity relation tend to be under sampled in continuum-based surveys as their spectra are typically dominated by emission from newly forming stars. We search for galaxies with high \OIII $\lambda 5007$ / \OII $\lambda 3727$, implying highly ionized nebular emission often indicative of low metallicity systems. With the Second Generation Low Resolution Spectrograph on the Hobby Eberly Telescope we acquired follow-up spectra, with higher resolution and broader wavelength coverage, of each low-metallicity candidate in order to confirm the redshift, measure the H$\alpha$ and \NII\ line strengths and, in many cases, obtain deeper spectra of the blue lines. We find our galaxies are consistent with the mass-metallicity relation of typical low mass galaxies. However, galaxies in our sample tend to have similar specific star formation rates (sSFRs) as the incredibly rare ``blueberry" galaxies found in \citet{yan17}. We illustrate the power of spectroscopic surveys for finding low mass and metallicity galaxies and reveal that we find a sample of galaxies that are a hybrid between the properties of typical dwarf galaxies and the more extreme blueberry galaxies.

%This pilot study predicts that a large sample of these objects will be identified in the complete HETDEX survey. 

\end{abstract}



\section{Introduction} 
\label{sec:intro}

Assembling large samples of galaxies in the low-redshift Universe has allowed astronomers to study properties of galaxies and how they recycle baryons in a statistical way. Wide-field photometric surveys such as the Sloan Digital Sky Survey (SDSS) \citep{aba09} have revealed much about the nature of galaxy evolution. For example, \citet{tre04} presented the mass-metallicity relation (MZR) with $\sim$53k galaxies from SDSS. This trend revealed that metal content tends to increase with mass for low mass galaxies and eventually flattens out at high mass \citep{tre04, ber12, and13}. Later studies reveal the fundamental metallicity relation (FMR) showing that star formation rate (SFR) causes scatter around the MZR \citep{man10, per13, and13}. \par

These statistical studies with large samples of galaxies uncovered clues as to how galaxies form stars and recycle materials. However, photometric surveys tend to under-sample the low-mass, low-metallicity populations as these galaxies tend to be faint and harder to pick up in broad band filters unless they are nearby and resolved or extreme in their emission line equivalent widths (EWs). Many studies have built samples of low-mass galaxies to better understand their nature and evolution.\par

Blue compact dwarfs (BCDs) were one of the first populations of low-redshift, low-mass galaxies discovered \citep{sar70}.  Since low-metallicity galaxies are expected to be dominated by young O and B stars, these galaxies are selected by their blue colors, compact morphologies, and low stellar masses \citep{sar70, kun00}. BCDs are strong UV emitters, suggesting far more active star formation than typical low-redshift galaxies \citep{lia16}. \par

More recently, \citet{kak07} discovered moderate-redshift strong emission line galaxies via narrow-band imaging in a survey meant to find high-redshift Lyman-alpha emitting (LAE) galaxies.  Subsequent spectroscopic observations from \citet{kak07} and \citet{hu09}, found 189 \OIII $\lambda$5007 emitters  with magnitude $<$ 24 in the two narrow filters covering 8091-8211~\AA\ and 9077-9198~\AA\ (\emph{z}=0.616-0.640 and \emph{z}=0.813-0.837 for \OIII $\lambda$5007 emitters). Of the 28 objects in their sample that had \OIII $\lambda$4363 detections, seven had log(O/H)+12 $<$ 7.65, which they defined as extremely metal poor galaxies (XMPGs). \par

As part of the Galaxy Zoo Project, \citet{car09} discovered a similar class of objects and named them ``green peas", due to their appearance in color images. These galaxies were selected with broad-band photometry to have very high \OIII $\lambda 5007$ equivalent widths (EW) (42-2388$\AA$) in the \emph{r}-band, limiting the redshift of the population to 0.112$<$\emph{z}$<$0.360. Green peas are also known for their compact morphologies, high specific star formation rates, and low masses. \citet{amo10} and \citet{haw12} both found green peas to have low metallicity for their mass. \citet{car09} found 251 of these galaxies in all 8423~deg$^2$ of the Sloan Digital Sky Survey (SDSS) data release (DR) 7 \citep{aba09}. \par

In an attempt to find the lowest redshift equivalent to green peas, \citet{yan17} searched through SDSS DR12 \citep{ala15} (14,555~deg$^2$) for galaxies with incredibly high \OIII $\lambda 5007$ EW (422-2635$\AA$), employing a similar color-color selection in bluer bands. They identified 43 ``blueberry galaxies'' at $z<0.05$ by their very blue colors and compact morphologies. These galaxies have extremely high [OIII] $\lambda 5007$ EW values and high specific star formation rates. Blueberries have low metallicities even for their low masses, and are even rarer than green peas. \par

\citet{ber12} estimated masses and metallicities for a sample (42 in sample, 19 with metallicity estimates) of local, typical dwarf galaxies from the Local Volume Legacy Sample (LVL) \citep{dal09}. They find an MZR at the low-mass end that agrees with the literature. \par

These studies selected these populations from photometric surveys, and subsequently observed them spectroscopically to confirm their detections and infer metallicities. Broad-band surveys like SDSS \citep{yor00} are able to search large areas of the sky and identify low-redshift galaxies via their red continuum emission. low-redshift galaxies that have low metallicity for their mass are likely faint continuum sources and can be easily confused with galaxies at higher redshift, posing a detection and classification challenge. Low-redshift, star-forming galaxies that currently have low metal content likely have spectra dominated by emission line regions associated with massive, young O and B stars. Since they will also tend to be low-mass they are likely faint in their red continuum from their older stellar populations. This prevents broad-band photometry from finding all but the very nearby resolved galaxies and the highest \OIII $\lambda$5007 equivalent width systems.  \par

Narrow-band imaging has the advantage of being able to efficiently identify faint galaxies based on their emission lines. However, this technique only works over the narrow redshift range of the filter. Thus, these surveys can detect  \OIII $\lambda$5007 to much fainter limits than broad band surveys, but are restricted to searching a much smaller volume. \par

There are only a small handful of galaxy samples assembled spectroscopically, without photometric pre-selection, that exist. One notable example is the KPNO International Spectroscopic Survey (KISS; \citealt{sal00}) . This objective-prism survey provides a sample of emission-line galaxies over two bandpasses: 6400-7200~\AA~(KISS red) and 4800-5500~\AA\ (KISS blue). From the KISS survey, \citet{bru20} built a sample of green pea-like galaxies by searching for \OIII $\lambda$5007 emission in their red prism covering \emph{z} = 0.29-0.41. A total of 15 of these 38 \OIII $\lambda$5007 emitting galaxies fall in the same parameter space as green peas in terms of their stellar mass, metallicity and SFR. \par

Spectroscopy is most sensitive to detecting emission line sources, regardless of how faint they are in the continuum, and typically has a broader bandpass allowing for a wider redshift range to be searched.  Slitless objective prism and grism surveys were the first to make large area spectroscopy possible; however, extracting the spectra of emission line galaxies from these data can be challenging in crowded fields, and the frames are background limited by their nature \citep{gal96, sal00, pir04, van11, bra12}. The sensitivity of objective-prism spectra is limited by the full sky background in the bandpass being incident on every pixel.\par 

Integral field spectrographs (IFS) overcome this challenge as they divide the field of view into spatial elements, each of which produces a spectrum. However, most IFS have small fields of view, so most studies focus on small areas or rely on pre-selection of targets \citep{all02, san12, bla13, bun15, bry15, bry16, mas18b}. Recently, however, the Hobby Ebery Telescope Dark Energy Experiment (HETDEX) has come online (Gebhardt et al., 2021, in preparation, Hill et al. 2021, in preparation). HETDEX utilizes Visible Integral-field Replaceable Unit Spectrographs (VIRUS; \citealt{hil18a}) to survey a $\sim$90~deg$^2$ area (a 1/4.5 fill factor within 540~deg$^2$ area) via blind spectroscopy, without target pre-selection. The main motivation for this survey is to measure the expansion history of the universe, and thereby probe the history of dark energy, via the large-scale structure of a sample of $\sim$1 million Lyman-$\alpha$ emitting galaxies in the redshift range $1.9<z<3.5$ (\citealt{hil08b,hil16}, Gebhardt et al., 2021, in preparation). HETDEX will also detect a similar number of \OII\ emitting galaxies at $z < 0.5$, and obtain a spectrum for any object that falls on the $\sim$35,000 fibers of the instrument. HETDEX can detect emission-lines as faint as $\sim 4 \times 10^{-17}$~ergs~cm$^{-2}$~s$^{-1}$. With much more light collecting area and its instrument's ability to detect galaxies from their emission alone, HETDEX can detect low-redshift star forming galaxies with much fainter continuum than those targeted for SDSS spectroscopy\null. It can therefore probe populations that are missing or underrepresented in magnitude-limited galaxy samples. \par

As a pathfinder for HETDEX, the HETDEX Pilot Survey (HPS) used the George and Cynthia Mitchell Spectrograph, the VIRUS Prototype instrument \citep{hil08a}, on the 2.7m Harlan J. Smith Telescope at McDonald Observatory to survey several patches of sky for emission-line galaxies without target pre-selection. \cite{ind19} constructed a complete sample of 29 \OII $\lambda 3727$ and \OIII $\lambda 5007$ emitting galaxies from the survey's 163~{arcmin}$^2$ area at $z<0.15$. This sample spanned a large mass range ($8.71< \log(M_{*}/M_{\odot})<10.80$) with fairly typical metallicities for their mass, though slightly higher than typical SFRs compared to SDSS star forming galaxies. However, two of the lowest mass galaxies in the sample have mass and metallicity values that fall in a similar regime as the green peas. The discovery of these two galaxies, without SDSS spectra, motivated our search through this first internal HETDEX data release for low mass and metallicity sources underrepresented in photometric surveys.  \par

The entire HETDEX survey will cover a volume over 600 times larger than HPS. We expect HETDEX to find a large sample of \OIII $\lambda 5007$ emitting galaxies that fill in the low mass and metallicity end of the MZR. The HETDEX survey is in its early stages, with observations of the spring field about 1/3 complete and the fall field about 1/4 complete. At this stage we have not attempted to build a complete sample of  \OIII $\lambda 5007$ emitting galaxies; rather, we use the early internal data release, HETDEX Data Release 1 (iHDR1), which contains $\sim9\%$ of the total HETDEX area, to search for candidates. We seek to find galaxies in the nearby Universe with high \OIII $\lambda 5007$/\OII $\lambda 3727$ ratios as these tend to be low metallicity. The spectra and fluxes presented for each object in our sample are from the most recent internal data release (iHDR2.1). This study demonstrates that even with conservative selection criteria these types of objects are more easily found with spectroscopy than in continuum-selection techniques. \par

Section~\ref{sec:data} describes the sample selection from an early HETDEX data release (Section~\ref{sec:hdr1}), spectra with the Second Generation Low Resolution Spectrograph (LRS2) (Section~\ref{sec:lrs2}), the LRS2 data processing (Section~\ref{sec:lrs2_redux}), and the photometric data gathered for spectral energy distribution (SED) fitting (Section~\ref{sec:photo}). The line fluxes measured from VIRUS and LRS2 spectra are discussed in Section~\ref{sec:fluxes}. Section~\ref{sec:prop} presents the properties determined for our sample of high  \OIII $\lambda 5007$/\OII $\lambda 3727$ (O3O2) ratio galaxies including their metallicities (Section~\ref{sec:met}), stellar masses (Section~\ref{sec:mass}), star formation rates (Section~\ref{sec:sfr}) and \OIII $\lambda 5007$ EWs (Section~\ref{sec:EW}). Section~\ref{sec:analysis} places this sample in context by comparing to  photometrically selected populations of galaxies in terms of the mass-metallicity relation (Section~\ref{sec:mass_met}) and the star forming galaxy sequence (Section~\ref{sec:mass_sfr}). Each object in our sample is described in Section~\ref{sec:obj_dis}. Our findings are discussed in Section~\ref{sec:disussion}, and the results of this study are summarized in Section~\ref{sec:conclusions}. \par

In this paper, we adopt a $\Lambda$CDM cosmology with $H_0 = 70$~km~s$^{-1}$~Mpc$^{-1}$, $\Omega_M=0.3$, and $\Omega_{\Lambda}=0.7$. References to metallicity refer to the gas-phase  oxygen abundance, which is generally given by its ratio to hydrogen, i.e., log(O/H)+12. \par

\section{Sample Selection and Data} 
\label{sec:data}

\begin{deluxetable*}{lccccccc}[ht!]
\tablecaption{Basic properties of objects in the sample.}
\tablehead{\colhead{ID} & \colhead{RA (J2000)} & \colhead{DEC (J2000)} & \colhead{$z$} & \colhead{VIRUS [OII]3727 flux} & \colhead{VIRUS H$\beta$ flux} & \colhead{VIRUS [OIII]4959 flux} & \colhead{VIRUS [OIII]5007 flux}}
\startdata
O3ELG1 & 31.0511 & -0.2286 & 0.0780 & $27.35\pm2.95$ & $26.41\pm3.66$ & $46.79\pm2.75$ & $135.58\pm3.02$ \\
O3ELG2 & 8.2826 & -0.1793 & 0.0780 & $126.94\pm6.15$ & $139.29\pm11.81$ & $255.79\pm23.77$ & $748.45\pm41.44$ \\
O3ELG3 & 6.0817 & -0.0720 & 0.0840 & $6.9\pm0.52$ & $10.39\pm1.55$ & $13.26\pm0.67$ & $36.88\pm1.67$ \\
O3ELG4a & 9.8204 & -0.0121 & 0.0820 & $54.76\pm2.94$ & $29.87\pm2.41$ & $46.56\pm1.23$ & $138.75\pm10.80$ \\
O3ELG4b & 9.8210 & -0.0111 & 0.0820 & $86.05\pm6.38$ & $27.03\pm1.29$ & $24.05\pm3.57$ & $64.59\pm11.57$ \\
O3ELG5 & 9.7674 & 0.0360 & 0.0830 & $30.47\pm1.93$ & $15.06\pm1.95$ & $29.44\pm1.69$ & $90.19\pm3.07$ \\
O3ELG6 & 13.4199 & 0.0403 & 0.0700 & $14.08\pm1.16$ & $13.17\pm0.74$ & $17.15\pm1.21$ & $54.91\pm1.96$ \\
O3ELG7 & 32.9842 & 0.0585 & 0.0960 & $21.15\pm2.00$ & $38.15\pm8.94$ & $90.61\pm15.12$ & $279.04\pm35.00$ \\
O3ELG8 & 31.8456 & 0.1995 & 0.0630 & $18.61\pm2.39$ & $12.76\pm1.79$ & $32.08\pm3.00$ & $88.86\pm5.98$ \\
O3ELG9 & 203.8739 & 51.0155 & 0.0620 & $707.58\pm79.32$ & $951.22\pm73.34$ & $1296.08\pm149.36$ & $4402.79\pm194.49$ \\
O3ELG10 & 197.6288 & 51.0607 & 0.0330 & $2.08\pm1.42$ & $3.3\pm0.40$ & $2.9\pm0.95$ & $11.0\pm0.84$ \\
O3ELG11 & 176.0732 & 51.1047 & 0.0840 & $93.06\pm20.99$ & $66.35\pm17.09$ & $156.72\pm26.15$ & $474.84\pm78.70$ \\
O3ELG12a & 200.1701 & 51.1910 & 0.0730 & $40.43\pm9.82$ & $43.6\pm7.51$ & $34.46\pm4.72$ & $113.35\pm8.73$ \\
O3ELG12b & 200.1696 & 51.1921 & 0.0730 & $18.71\pm0.73$ & $34.94\pm3.50$ & $53.08\pm3.14$ & $163.0\pm3.07$ \\
O3ELG13 & 172.8717 & 51.2003 & 0.0650 & $180.2\pm28.67$ & $116.19\pm5.48$ & $204.44\pm9.14$ & $509.39\pm34.44$ \\
O3ELG14 & 176.4366 & 51.2577 & 0.0900 & $93.13\pm4.63$ & $103.72\pm3.23$ & $152.1\pm3.55$ & $302.33\pm44.20$ \\
O3ELG15 & 212.7970 & 51.2664 & 0.0270 & $199.33\pm13.48$ & $140.94\pm22.85$ & $235.24\pm18.38$ & $646.68\pm112.73$ \\
O3ELG16 & 168.1299 & 51.8901 & 0.0710 & $9.05\pm2.47$ & $15.6\pm2.15$ & $21.2\pm2.31$ & $60.22\pm5.04$
\enddata
\tablecomments{Flux units are $10^{-17}$~ergs~s$^{-1}$~cm$^{-2}$. A description of the VIRUS flux measurements is given in Section \ref{sec:fluxes}. *Note O3ELG4b is not included in our sample as it did not meet our selection criteria but its properties are presented since it is a companion of O3ELG4a. \label{samp_tbl}}
\end{deluxetable*}

\begin{figure*}[ht!]
\epsscale{1.0}
\plotone{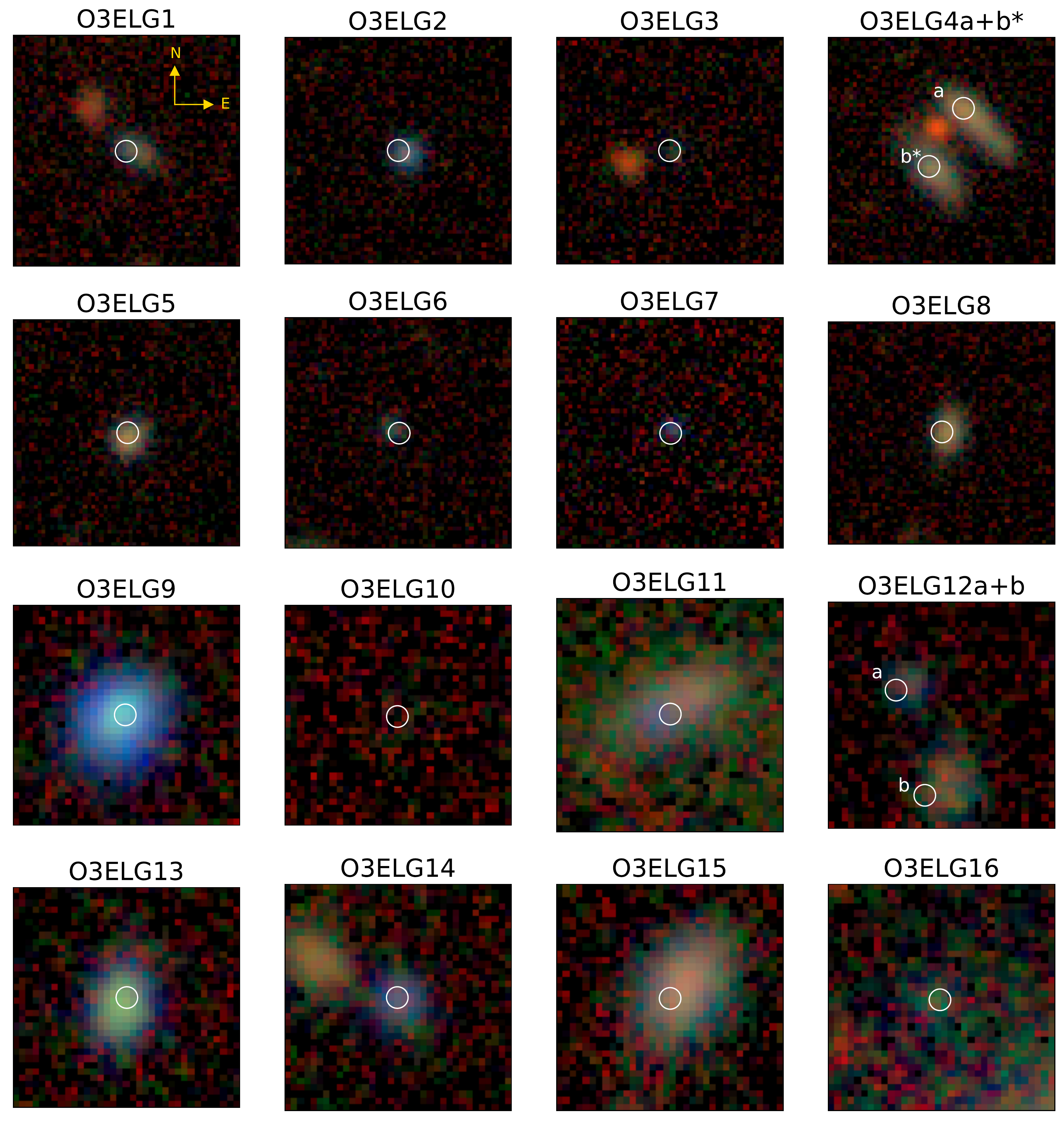}
\caption{Legacy Survey images in $g$, $r$, and $z$ for the 17 galaxies in our sample, and also O3ELG4b. Each cutout is $14\arcsec$ on a side, and is labeled with its object name. The orientation is shown in the first panel. The centroid of the object are indicated with a white circle. O3ELG4 and O3ELG12 have two sources at the same redshift and the double sources are represented in just one panel. For O3ELG12 both sources are included in the sample. For O3ELG4 only source a is included in our sample. O3ELG4b is found to be at the same redshift at O3ELG4a but does not meet the search criteria for our sample. For O3ELG4 and O3ELG12 the image is centered on the HETDEX detection coordinate and there are two white circles representing each objects' central coordinates.}.\label{fig:cutouts}
\end{figure*}

%$\OIII\lambda$5007/$\OII\lambda$3727

We constructed our sample of low-metallicity \OIII\ emitters from an early internal HETDEX data release (iHDR1) consisting of a catalog of every emission line detection in the 8.16~deg$^2$ of the survey completed between January 3rd, 2017 and February 9th, 2019. The HETDEX survey is conducted with the Visible Integral Field Unit Spectrograph (VIRUS) which covers 3500-5500~\AA\ at an $R\sim800$ (2 \AA/pixel dispersion). VIRUS will consist of 78 spectrograph units each fed a square bundle of 448 fibers. Each fiber bundle covers $\sim$50"x50" area field with 1.5" fibers. Each shot for the survey is covered by 3 exposures with small offsets (referred to as dithers) to fill in the gaps between the fibers within each bundle's field. The survey is able to detect \OIII $\lambda 5007$ below $z=0.1$ and above a monochromatic flux limit of $\sim 4 \times 10^{-17}$~ergs~cm$^{-2}$~s$^{-1}$ (above 5$\sigma$; Hill et al. 2021, in prep.). Spectra for each object were re-extracted from the most recent data release, iHDR2.1. VIRUS data are reduced with the an automatic pipeline discussed in Gebhardt et. al. (in prep. 2021). 

Each galaxy in our sample was followed-up with the Second Generation Low Resolution Spectrograph (LRS2) Red unit to obtain lines outside the VIRUS bandpass to obtain better metallicity fits, H$\alpha$ SFRs, and, in some cases, to confirm objects redshifts. Most candidates were additionally followed up with LRS2 Blue unit observations to obtain deeper and more accurate line fluxes. The LRS2 observations are discussed in detail in Section~\ref{sec:lrs2}. Photometric data were collected from several surveys to perform SED fitting of our sample. These data are discussed in detail in Section~\ref{sec:photo}. The basic properties of our final sample can be found in Table~\ref{samp_tbl}.

\subsection{Sample Selection from HETDEX} 
\label{sec:hdr1}

Candidates for our sample were first selected from an early internal data release of the HETDEX emission-line catalog. We selected objects initially based on \OIII $\lambda 5007$, and made cuts to require high \OIII $\lambda 5007$/\OII$\lambda 3727$ since this high ionization ratio generally implies low metallicity. The goal was to survey the early HETDEX data for very low-metallicity candidates, not to try to build a complete sample. To find candidate  \OIII $\lambda 5007$ emitters, we searched the HETDEX catalog for emission line detections redward of 4980~\AA . These \OIII $\lambda 5007$ detections were kept in the sample if the Gaussian fit to the line was detected with signal-to-noise ratio (S/N) $>$ 10 and a $\chi^{2}<$ 6.0. The high S/N cut was chosen to allow for the detection of  \OIII $\lambda 4959$, as this feature would confirm the line identification. 
For the \OIII $\lambda 5007$ line detection candidates that met these criteria, spectra were extracted around each detection from the VIRUS data using a point spread function (PSF) weighted extraction. Objects that fall near the edge of an integral field unit (IFU) are susceptible to having a fraction of their PSF shift out of the fibers due to atmospheric differential refraction (ADR\null). A wavelength resolution element was flagged if it had less than 70\% coverage from its weighted fibers. If more than half of the spectrum was flagged it was rejected from the sample. We also removed spectra from VIRUS observations (made up of three dithered exposures) with seeing worse than $3\farcs 5$ or with system (HET combined with VIRUS) throughputs below 1\% at 4540~\AA\ through that fiber. These cuts left $\sim$25k candidates, however, many of these detections came from the same duplicate sources.  Finally, only candidate spectra with  \OIII $\lambda 4959$ fit at one third of the flux of  \OIII $\lambda 5007$ within the appropriate margin of error were kept in the sample, leaving 2361 detections. The \OIII $\lambda 5007$ and 4959 lines were fit with two Gaussians simultaneously with a fixed line width.  \par 

Candidates with 20 or more independent detections were removed as they were spatially extended, low-redshift galaxies. After this preliminary culling, only objects with \OIII $\lambda 5007$/\OII $\lambda 3727$ (O3O2)  $>$ 3.0 were kept in the sample as this high ratio implies low metallicity, leaving 444 candidate spectra. Four of us visually examined each candidate's spectrum and labeled the number of lines that were clearly identified in the spectra and with an O3O2 $>$ 5.0.  In total, 32 ``strong'' candidates with three or more emission lines were identified, along with 5 more ``moderate'' candidates  where there was considerable uncertainly on the identification of the lines. All of these candidates also have O3O2 $>$ 5.0. Many of the remaining 444 spectra were not included in our strong candidate list because a clear detection of  \OII $\lambda 3727$ or H$\beta$ could not be identified. These galaxies could be candidates with even lower metallicity, but with limited ability to obtain additional LRS2 data for confirmation, we chose to move forward with the most secure candidates in our sample. \par 

Since the candidates are relatively low-redshift sources, we suspected that many of the objects could be extended and hence have multiple VIRUS detections per source.  We therefore inspected the detections of the 32 strong candidates, and merged those objects that had identical redshifts and spatial separations less than $7\farcs 2$.  This reduced the number of unique strong candidates to 15.   \par

Since many of our sources have multiple detections in the first version of the HETDEX catalog, a mean RA and DEC were determined by the following method.  A $14\arcsec \times 14\arcsec$ map of  \OIII $\lambda 5007$ was built around the VIRUS detection(s) from the VIRUS IFU data (see Appendix~\ref{fig:spec_p1}). The centroid of the  \OIII $\lambda 5007$ in these maps defined the final coordinates of the source; these RA and DEC values are presented in Table~\ref{samp_tbl}. Final VIRUS spectra for each source were then constructed using a fixed window (3") optimal extraction based upon that coordinate. Final extracted spectra often show an O3O2 ratio smaller than the initial threshold of five. Final VIRUS spectra are presented in Figures ~\ref{fig:spec_p1} and ~\ref{fig:spec_p2} in the appendix.\par

These objects were found by searching through iHDR1. However, VIRUS spectra for each object were re-extracted using the most recent internal data release, iHDR2.1. Details of the data reduction pipeline and the flux calibration can be found in Gebhardt et. al. (2021 in prep.). We correct the spectra for of galactic extinction along their line-or-sight based on galactic reddening estimates from \citet{sch11}. \par

\subsection{LRS2 Follow-up} 
\label{sec:lrs2}

We followed up the 15 strong and three moderate candidates with the LRS2 (\citealt{cho16}, Hill et al. 2021, in prep.), an integral field spectrograph on the Hobby-Eberly Telescope at McDonald Observatory. LRS2 is comprised of two dual-beam spectrograph units: LRS2 Blue (LRS2-B) with UV (R$\sim$2474, 0.496$\AA$/pixel dispersion) and orange (R$\sim$1477, 1.192 \AA/pixel dispersion) channels covering 3700-4700~\AA\ and 4600-7000~\AA, respectively, and LRS2 Red (LRS2-R) with red (R$\sim$2480, 0.978 \AA/pixel dispersion) and far-red (R$\sim$2561, 1.129$\AA$/pixel dispersion) channels covering 6500-8420~\AA\ and 8180-10500~\AA, respectively.  Each unit covers a $6\arcsec \times 12\arcsec$ field with full fill factor due to microlens array coupling to the fiber feed
with $0\farcs 59$ lenslet pitch and 170~$\micron$ fiber core diameter  \citep{cho16}. The average seeing of our observations was $1\farcs 8$. 

We were able to obtain LRS2-R data for all 15 strong candidates, and LRS2-B data on 14 of the them. LRS2-B data were principally used to better measure  \OII $\lambda 3727$, which is typically weakly detected in the VIRUS spectra. The LRS2 observations of O3ELG4 and O3ELG12 revealed two spatially separate  \OIII $\lambda 5007$ sources at the same redshift. We refer to these detections as individual sources O3ELG4a and b and O3ELG12a and b. The HETDEX detection coordinates are centered between O3ELG12a and b, both meet our selection criteria, and are included as two separate objects in our sample. The O3ELG4 HETDEX detection was centered on O3ELG4a. We extracted VIRUS and LRS2 spectra of O3ELG4b and determined its properties since this object was of interest as it is a companion of O3ELG4a. However, O3ELG4b does not meet the selection criteria of this sample and its determined properties reveal it is to be a typical field galaxy (properties are discussed in Section~\ref{sec:prop}). For this reason O3ELG4b is not included as part of our sample, however, its properties are presented in paper tables.  \par

\begin{figure*}
\begin{center}
\includegraphics[width=.9\textwidth]{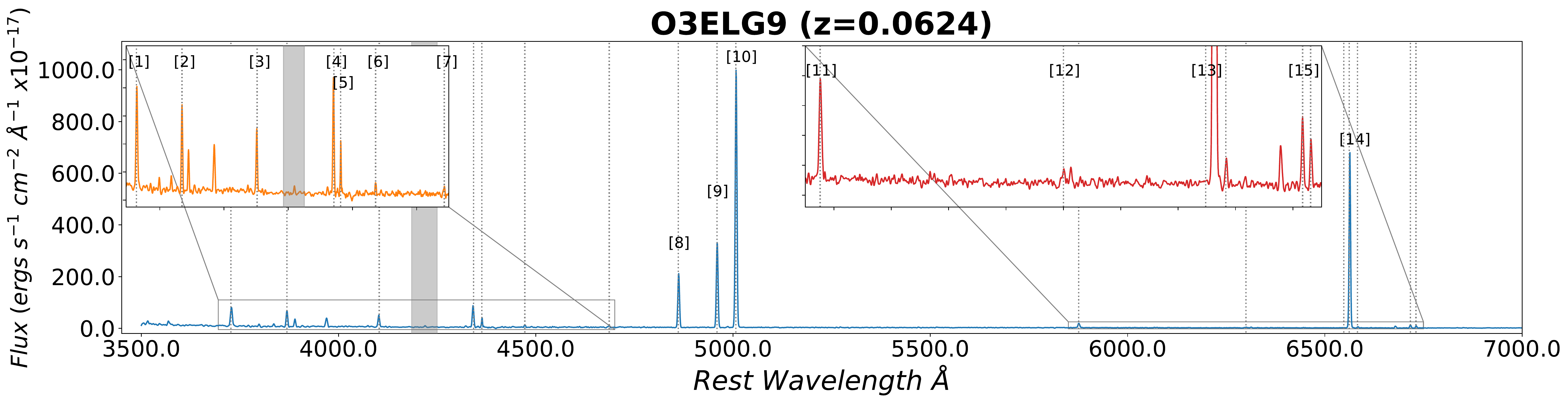}
\includegraphics[width=.9\textwidth]{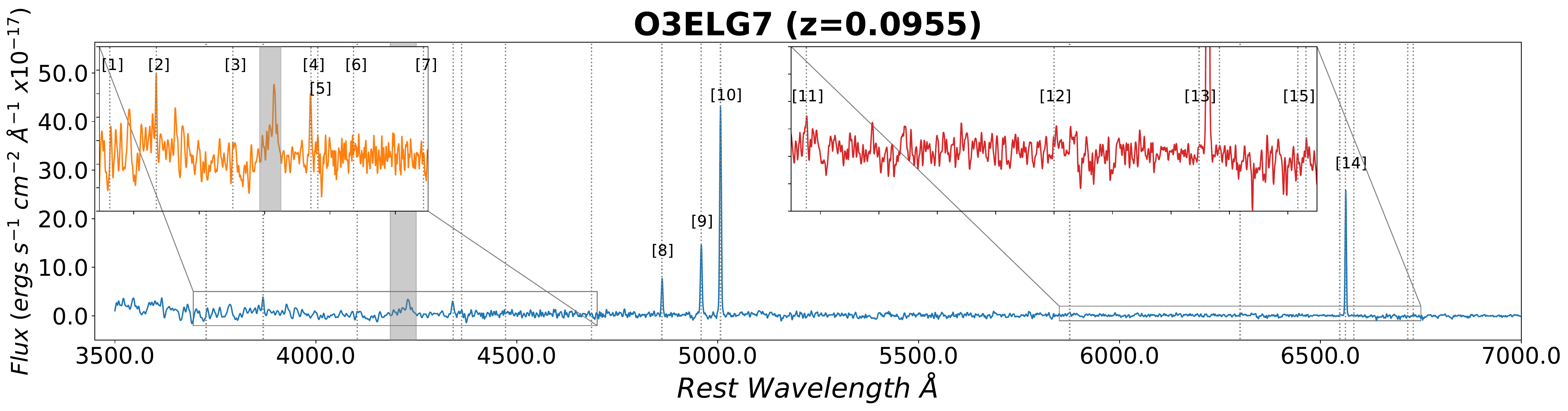}
\includegraphics[width=.9\textwidth]{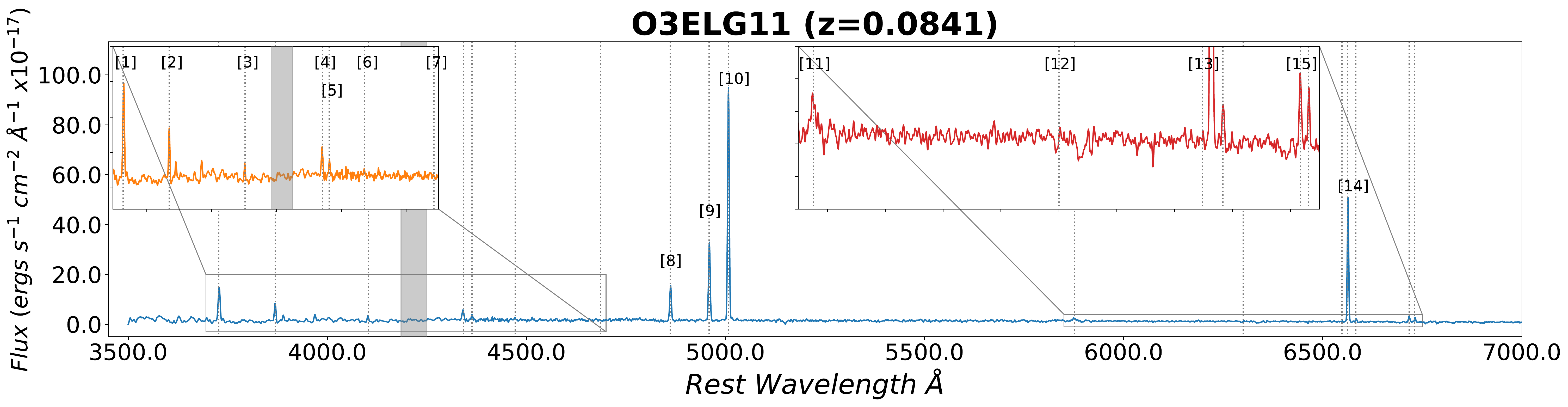}
\end{center}
\caption{Three examples of our LRS2-B and LRS2-R spectra.  The measured emission lines are highlighted with grey dotted lines. The transition between the LRS2-B UV and orange channel shows a feature associated with the dichroic filter around 4200~\AA\null. This region is masked in grey. The top spectrum is from our brightest object, where many faint lines are clearly measured. The middle spectrum is from the object with the highest  \OIII $\lambda 5007$/\OII $\lambda 3727$ ratio. The third spectrum is from another bright object in our sample with the \OIII $\lambda 4363$ line detected at greater than 5$\sigma$ confidence. Labeled emission lines: [1] [O II]$\lambda$3727, [2] [Ne III]$\lambda$3869, [3] H$\delta$, [4] H$\gamma$, [5] [O III]$\lambda$4363, [6] He I $\lambda$4471, [7] He II $\lambda$4686,[8] H$\beta$, [9] [O III]$\lambda$4959, [10] [O III]$\lambda$5007, [11] He I $\lambda$5876, [12] [O I]$\lambda$6300, [13] [N II]$\lambda\lambda$6548,6583, [14] H$\alpha$, [15] [S II]$\lambda\lambda$6717,6731}
\label{fig:lrs2_spec}
\end{figure*}

We were additionally able to follow up three of our moderate candidates. One of these sources (O3ELG10) was confirmed to be an \OIII $\lambda 5007$ detection while the other two were higher redshift \OII $\lambda 3727$. Combined with the 16 strong sources (which includes O3ELG4a and O3ELG12a+b), we have 17 objects in our final sample. Details for individual LRS2 observations can be found in Table~\ref{lrs2_tbl}, and example spectra are shown in Figure~\ref{fig:lrs2_spec}.\par

Since VIRUS data are limited to the 3500-5500~\AA\ bandpass, the LRS2 additional spectrum of each object added information for the derivations of metallicity and SFR. Although multiple lines were identified in candidate VIRUS spectra, at a \OIII $\lambda 5007$ SNR=10, the SNR of \OIII $\lambda 4959$ is often only a $\sim$3$\sigma$ detection. For a few of our faint sources, this additional follow up was needed to confirm that the faint lines were not correlated noise spikes. Measurement of H$\alpha$, in particular, confirmed the redshift and line identifications.\par

With all strong candidates proven to be actual sources by the LRS2 follow-up, we are confident that visual identification of at least three emission lines in VIRUS data is an effective way for finding these sources. In future studies with the larger HETDEX sample, where follow-up of every source is not feasible, we can rely on this selection method. \par

\subsection{LRS2 Reduction}
\label{sec:lrs2_redux}

We used the HET pipeline, Panacea\footnote{https://github.com/grzeimann/Panacea}, to analyze and reduce our LRS2 observations.  The pipeline produces an array of wavelength and flux-calibrated fiber spectra.  We designed a custom script to use those spectra to model each source's 2D spatial profile using bright emission lines, subtract the sky emission using fibers $>5\arcsec$ from the source, and optimally extract the source using our 2D spatial profile, taking into account the differential atmospheric refraction as function of wavelength.  Reduction was performed separately for the two LRS2 units, LRS2-B and LRS2-R\null. The two spectra for a given source were combined using the overlap region of the LRS2-B orange and LRS2-R red channels, normalizing the spectra to the LRS2-B orange channel flux.

\begin{deluxetable}{lccc}[ht!]
\tablecaption{LRS2 observations for follow up of the 16 out of the 17 objects in our sample.  \label{lrs2_tbl}}
\tablehead{\colhead{ID} & \colhead{Observation Date} & \colhead{Channel} & \colhead{Exposure Time (s)}}
\startdata
O3ELG1 & 2019-10-30 & B & 1200 \\
  & 2019-10-30 & R & 1200 \\
O3ELG2 & 2019-12-22 & B & 900 \\
  & 2019-12-22 & R & 900 \\
O3ELG3 & 2019-12-21 & B & 900 \\
  & 2019-12-21 & R & 900 \\
O3ELG4a+b* & 2020-07-28 & B & 900 \\
  & 2020-07-28 & R & 900 \\
O3ELG5 & 2019-12-19 & B & 900 \\
  & 2019-12-19 & R & 900 \\
O3ELG6 & 2019-12-16 & B & 900 \\
  & 2019-12-16 & R & 900 \\
O3ELG7 & 2019-10-29 & B & 1200 \\
  & 2019-10-29 & R & 1200 \\
O3ELG8 & 2019-11-27 & R & 1200 \\
O3ELG9 & 2020-04-15 & B & 600 \\
  & 2020-04-15 & R & 600 \\
O3ELG10 & 2020-04-13,27 & B & 2400 \\
  & 2020-03-22 & R & 1200 \\
O3ELG11 & 2019-12-21 & B & 900 \\
  & 2019-12-21 & R & 900 \\
O3ELG12a+b & 2020-03-22 & B & 900 \\
  & 2020-03-22 & R & 900 \\
O3ELG13 & 2020-05-27 & B & 1200 \\
  & 2020-03-24 & R & 1800* \\
O3ELG14 & 2020-05-10 & B & 1200 \\
  & 2020-03-25 & R & 900 \\
O3ELG15 & 2020-04-13 & B & 900 \\
  & 2020-04-13 & R & 900 \\
O3ELG16 & 2020-05-29 & B & 1800 \\
  & 2020-03-31 & R & 900 \\
\enddata
\tablecomments{For the LRS2-R observation of O3ELG13 the exposure time was doubled to account for poor conditions. *Note O3ELG4a was the object found from our selection criteria but the LRS2 observations revealed a companion object at the same redshift, O3ELG4b. However, this O3ELG4b is not included in our sample count as it does not meet our selection criteria.}
\end{deluxetable}

\nopagebreak

\subsection{Photometric Data}
\label{sec:photo}

Photometric data were gathered for spectral energy distribution (SED) fitting from The DESI Legacy Imaging Surveys \citep{dey19}, The Dark Energy Survey \citep[DES;][]{abb18}, and the Galaxy Evolution Explorer \citep[GALEX;][]{mar05}\null. Despite these objects all being at low redshift, they are fairly faint, so we required the deepest available wide field catalogs. The DESI Legacy Imaging Surveys (LS) provide optical bands $g$, $r$, and $z$: The Dark Energy Camera Legacy Survey (DECaLS) covered the HETDEX spring field objects, while the Beijing-Arizona Sky Survey (BASS) and The Mayall $z$-band Legacy Survey (MzLS) covered the HETDEX spring field. Four Wide-field Infrared Survey Explorer \citep[WISE;][] {wri10}
%(WISE)\citep{wri10} 
bands (W1-4) from AllWISE were also provided as part of the LS for all objects. These galaxies range in $g$ magnitudes from 18.9 to 23.3. Photometric bands from the LS came from the data release 8 main tractor catalog. \par

\begin{splitdeluxetable*}{lcccccccBcccccccc}
\tablecaption{Photometry from all surveys presented in Table~\ref{phot_tbl}. \label{phot_bands_tbl}}
\tablehead{\colhead{ID} & \colhead{LS mag $g$} & \colhead{LS flux $g$} & \colhead{LS flux $r$} & \colhead{LS flux $z$} & \colhead{LS flux w1} & \colhead{LS flux w2} & \colhead{LS flux w3} & \colhead{LS flux w4} & \colhead{DES flux $g$} & \colhead{DES flux $r$} & \colhead{DES flux $i$} & \colhead{DES flux $z$} & \colhead{DES flux $Y$} & \colhead{GLX flux fuv} & \colhead{GLX flux nuv}}
\startdata
O3ELG1 & 20.30 & $7.59\pm0.10$ & $7.9\pm0.12$ & $8.22\pm0.34$ & $0.98\pm2.31$ & $6.16\pm5.26$ & --- & $864.78\pm874.04$ & $7.09\pm0.20$ & $7.31\pm0.27$ & $8.18\pm0.45$ & $9.09\pm0.82$ & $11.92\pm2.51$ & $1.77\pm0.35$ & $3.15\pm0.54$ \\
O3ELG2 & 20.15 & $8.73\pm0.08$ & $6.83\pm0.10$ & $4.64\pm0.20$ & $2.67\pm2.25$ & --- & $245.42\pm135.64$ & --- & $8.34\pm0.19$ & $6.41\pm0.32$ & $8.08\pm0.49$ & $5.51\pm0.74$ & $6.84\pm2.85$ & $2.97\pm0.29$ & $3.51\pm0.52$ \\
O3ELG3 & 21.86 & $1.8\pm0.10$ & $1.71\pm0.10$ & $0.93\pm0.33$ & $5.14\pm2.22$ & $6.19\pm5.27$ & $35.56\pm136.01$ & --- & $1.91\pm0.20$ & $2.3\pm0.24$ & $1.65\pm0.49$ & $1.04\pm0.88$ & $3.57\pm3.01$ & $0.77\pm0.24$ & $1.13\pm0.41$ \\
O3ELG4a & 19.27 & $19.67\pm0.14$ & $23.62\pm0.19$ & $26.16\pm0.44$ & $11.55\pm2.40$ & --- & $363.33\pm140.73$ & $9061.21\pm1169.06$ & $19.69\pm0.29$ & $24.68\pm0.52$ & $27.0\pm0.61$ & $28.9\pm1.13$ & $26.73\pm3.77$ & --- & $6.17\pm2.84$ \\
O3ELG4b* & 19.01 & $24.85\pm0.15$ & $33.29\pm0.20$ & $40.06\pm0.47$ & $22.8\pm2.47$ & $10.6\pm5.70$ & $128.35\pm144.19$ & $14003.17\pm1182.41$ & $23.48\pm0.27$ & $31.33\pm0.52$ & $39.1\pm0.62$ & $38.99\pm1.12$ & $45.55\pm3.77$ & $8.43\pm0.82$ & $10.92\pm0.82$ \\
O3ELG5 & 19.91 & $10.91\pm0.09$ & $13.21\pm0.12$ & $15.41\pm0.28$ & $13.96\pm2.29$ & $10.95\pm5.33$ & $35.88\pm145.82$ & $356.7\pm1217.26$ & $9.71\pm0.16$ & $12.83\pm0.33$ & $15.04\pm0.39$ & $14.43\pm0.71$ & $8.88\pm2.37$ & $2.82\pm0.51$ & $1.95\pm0.52$ \\
O3ELG6 & 21.48 & $2.57\pm0.08$ & $2.57\pm0.09$ & $2.21\pm0.27$ & $0.06\pm2.33$ & $5.77\pm5.44$ & $27.18\pm184.90$ & $333.23\pm1289.10$ & $2.63\pm0.23$ & $2.58\pm0.33$ & $2.23\pm0.50$ & $2.13\pm0.90$ & $0.81\pm2.97$ & --- & --- \\
O3ELG7 & 21.29 & $3.05\pm0.10$ & $1.84\pm0.12$ & $1.54\pm0.32$ & --- & --- & $36.21\pm97.27$ & --- & $2.52\pm0.14$ & $1.55\pm0.30$ & $2.81\pm0.47$ & $0.25\pm0.61$ & --- & $1.17\pm0.39$ & --- \\
O3ELG8 & 19.95 & $10.47\pm0.09$ & $13.55\pm0.11$ & $15.88\pm0.32$ & $5.89\pm2.21$ & $3.6\pm5.11$ & $265.01\pm100.35$ & $2479.63\pm863.27$ & $9.67\pm0.18$ & $13.14\pm0.24$ & $14.62\pm0.35$ & $14.86\pm0.72$ & $16.93\pm2.28$ & $1.18\pm0.51$ & $1.81\pm0.78$ \\
O3ELG9 & 17.47 & $102.48\pm0.34$ & $73.29\pm0.43$ & $42.96\pm0.46$ & $26.7\pm1.71$ & $40.29\pm3.92$ & $464.59\pm95.98$ & $3906.85\pm880.41$ & --- & --- & --- & --- & --- & $56.26\pm9.97$ & $55.83\pm5.70$ \\
O3ELG10 & 21.90 & $1.74\pm0.28$ & $2.37\pm0.41$ & $1.5\pm0.90$ & --- & --- & --- & --- & --- & --- & --- & --- & --- & --- & --- \\
O3ELG11 & 19.76 & $12.53\pm0.17$ & $12.6\pm0.25$ & $13.33\pm0.36$ & $31.77\pm1.80$ & $4.93\pm4.06$ & $87.34\pm96.61$ & $1837.33\pm869.50$ & --- & --- & --- & --- & --- & $4.64\pm1.25$ & $18.63\pm1.66$ \\
O3ELG12a & 20.34 & $7.28\pm0.20$ & $9.43\pm0.31$ & $9.14\pm0.50$ & $0.96\pm1.60$ & --- & --- & $476.99\pm799.63$ & --- & --- & --- & --- & --- & --- & $3.49\pm1.28$ \\
O3ELG12b & 21.03 & $3.87\pm0.14$ & $3.12\pm0.20$ & $2.23\pm0.26$ & $1.99\pm1.55$ & $4.37\pm3.55$ & --- & $685.34\pm797.26$ & --- & --- & --- & --- & --- & --- & $8.62\pm1.97$ \\
O3ELG13 & 18.66 & $34.37\pm0.26$ & $35.3\pm0.38$ & $31.02\pm0.52$ & $18.65\pm1.79$ & $15.29\pm4.06$ & $76.52\pm99.82$ & --- & --- & --- & --- & --- & --- & $9.72\pm2.31$ & $11.48\pm1.31$ \\
O3ELG14 & 19.81 & $11.93\pm0.17$ & $7.87\pm0.26$ & $7.58\pm0.39$ & --- & $13.08\pm4.07$ & $107.55\pm100.66$ & --- & --- & --- & --- & --- & --- & $4.73\pm2.42$ & --- \\
O3ELG15 & 18.38 & $44.52\pm0.28$ & $49.72\pm0.45$ & $53.95\pm0.71$ & $25.87\pm1.52$ & $12.64\pm3.49$ & $108.03\pm80.71$ & --- & --- & --- & --- & --- & --- & --- & $7.74\pm1.28$ \\
O3ELG16 & 21.28 & $3.06\pm0.16$ & $2.22\pm0.28$ & $1.28\pm0.40$ & --- & --- & $82.27\pm105.21$ & $1885.67\pm956.74$ & --- & --- & --- & --- & --- & --- & --- \\
\enddata
\tablecomments{Photometry from the DESI Legacy Survey Imaging Surveys (LS) containing $g$, $r$, and $z$ bands from DESI and the WISE w1, w2, w3, and w4 infrared bands, the Dark Energy Survey (DES) containing $g$, $r$, $i$, $z$, and $Y$ bands for our fall field targets, and GALEX (GLX) containing fuv and nuv bands. All flux values are in microJanskys. *Note O3ELG4b is not included in our sample as it did not meet our selection criteria but its properties are presented since it is a companion of O3ELG4a. \label{phot_tbl}}
\end{splitdeluxetable*}

For objects in the fall field, additional optical data were available from the DES; DES reaches deeper than DECaLS and provided two additional bands, $i$ and $Y$. Photometric bands from DES came from the DES data release 1 main catalog. Both the LS and DES catalogs were queried through Astro Data Lab\footnote{http://datalab.noao.edu}.   \par

A GALEX search from the GALEX Release 6 and 7 was performed for all objects in our sample via MultiMission Archive at Space Telescope Science Institute (MAST). 12 objects returned matches from both the All Sky Imaging Survey (AIS) and Medium Imaging Survey (MIS). GALEX provides two UV bands, FUV and NUV, with depths depending on the survey match. Note for instruments with a large PSF like WISE and GALEX, there was not an additional effort to deblend sources. Due to this sources with a neighbor within 10 arcseconds (O3ELG1,3,14 and the double sources 4 and 12) may have some contamination in their WISE and GALEX bands. Fluxes for Legacy Survey, DES and GALEX bands are listed in Table~\ref{phot_bands_tbl}, as are the descriptions of the surveys and their depths.

\begin{deluxetable*}{cccccc}[ht!]
\tablecaption{Description of photometric surveys}
\tablehead{\colhead{Field} & \colhead{Filter} & \colhead{Survey} & \colhead{Telescope} & \colhead{Instrument} & \colhead{depth(AB)}}
\startdata
Spring Field & $g$ & Legacy Survey (DECaLS) & BLANCO & DECam & 23.72\\
     & $r$ & Legacy Survey (DECaLS) & BLANCO & DECam & 23.27\\
     & $z$ & Legacy Survey (DECaLS) & BLANCO & DECam & 22.22\\
    & W1(3.4$\micron$) & Legacy Survey (ALLWISE) & WISE & DESI & 20.0\\
     & W2(4.6$\micron$) & Legacy Survey (ALLWISE) & WISE & DESI & 19.3\\
     & W3(12$\micron$) &  Legacy Survey (ALLWISE) & WISE & DESI & variable \\
     & W4(22$\micron$) & Legacy Survey (ALLWISE) & WISE & DESI & variable \\
    & fuv & GALEX (AIS, MIS) & GALEX & FUV detector & 19.9,22.6\\
    & nuv & GALEX (AIS, MIS) & GALEX & NUV detector & 20.8,22.7\\
\hline
Fall Field & $g$ & Legacy Survey (BASS) & Bok & 90Prime & 23.48\\
    & $r$ & Legacy Survey (BASS) & Bok & 90Prime & 22.87\\
    & $z$ & Legacy Survey (MzLS) & Mayall & Mosaic-3 & 22.29\\
    & W1(3.4$\micron$) & Legacy Survey (ALLWISE) & WISE & DESI & 20.0\\
     & W2(4.6$\micron$) & Legacy Survey (ALLWISE) & WISE & DESI & 19.3\\
     & W3(12$\micron$) &  Legacy Survey (ALLWISE) & WISE & DESI &  variable\\
     & W4(22$\micron$) & Legacy Survey (ALLWISE) & WISE & DESI & variable \\
    &  $g$ & Dark Energy Survey &BLANCO & DECam & 24.45\\
    & $r$ & Dark Energy Survey & BLANCO & DECam & 24.3\\
    & $i$ & Dark Energy Survey & BLANCO & DECam & 23.5\\
    & $z$ & Dark Energy Survey & BLANCO & DECam & 22.90\\
    & $Y$ & Dark Energy Survey & BLANCO & DECam & 21.70\\
    & fuv & GALEX (AIS, MIS) & GALEX & FUV detector & 19.9,22.6\\
    & nuv & GALEX (AIS, MIS) & GALEX & NUV detector & 20.8,22.7\\
\enddata
\end{deluxetable*}

\section{Line Flux Measurements} 
\label{sec:fluxes}

We measured emission lines strengths from both the VIRUS and LRS2 spectra. Refer to Table~\ref{flux_pri_tbl} and \ref{flux_sec_tbl} for the full list of emission lines measured from each galaxy's LRS2 spectrum. All lines blueward of $5500/(1+z)\AA $ were also measured in each galaxy's VIRUS spectrum. For each spectrum, we constructed an underlying continuum model by masking the emission lines and smoothing the spectrum using a Gaussian kernel with a standard deviation of 8 pixels. We subtracted the continuum model and fit model Gaussians to each emission line using \texttt{Astropy's} modeling subpackage. Fluxes are estimated by integrating the fit Gaussian. Balmer lines are corrected for absorption measured from the best fit SED (refer to Section \ref{sec:mass} for details on the SED fitting). \par

Errors to the lines fits are estimated by using the expected variance of the $\chi^2$ distribution. We feel this is a fair approximation since there is only one degree of freedom in our fitted parameters, the flux of the Gaussian. For this approximation we assume homoscedastic errors. If we assume 
\begin{equation}
 \chi^2=\sum{\frac{{(a-b)}^2}{{\sigma}^2}}~\sim~N 
\end{equation}
where a is the measured, continuum-subtracted, emission line spectrum, b is the Gaussian fit, and N is the number of degrees of freedom, we can solve for $\sigma$. Here sigma represents the error for each wavelength of the spectrum. We approximate the fit error by scaling this by the standard deviation of $\chi^2$ which can be approximated by $\sqrt{2*N}$. The fit errors are then approximated by the following equation:
\begin{equation}
 {error}_{\rm fit}~\approx~\sqrt{2*N}*\sqrt{\sum{\frac{{(a-b)}^2}{N}}}
\end{equation}

\par

%We subtracted the continuum model and summed the flux within $\pm 5$~\AA\ of the line of interest.  To estimate the error, we used the standard deviation of the continuum wavelengths between $7~{\rm\AA} < \lambda <  25~{\rm\AA}$ from the emission line added the error in quadrature for the number of summed pixels in our line flux measurement. 

Fluxes measured from LRS2 spectra were normalized to the VIRUS spectra; the scaling was based on the  \OIII $\lambda 5007$ LRS2 and VIRUS measurements. Since considerable efforts have been made to accurately flux calibrate VIRUS for the HETDEX survey (Gebhardt et. al. 2021\ in prep.) we choose to scale LRS2 to VIRUS\null. In the following analysis when line fluxes blueward of 5500~\AA\ are used the VIRUS fluxes are chosen. \par

Table~\ref{samp_tbl} contains flux values for some of the emission lines measured from VIRUS spectra, and Table~\ref{flux_pri_tbl} and \ref{flux_sec_tbl} provide flux values measured from LRS2\null. Since the LRS2 spectra are normalized to the VIRUS spectra based on the  \OIII $\lambda 5007$ flux, that line is not represented in the LRS2 flux tables. As a check of the line fluxes and their errors, we measure the \OIII $\lambda 5007$/\OIII $\lambda 4959$ ratio which has a known ratio of 2.98. This flux ratio of these measured lines agree within their error bars.

%\begin{turnpage}
\begin{deluxetable*}{lcccccccc}[ht!]
\tablecaption{LRS2 flux measurements for the lines primarily used in this study.  \label{flux_pri_tbl}}
\tablehead{\colhead{ID} & \colhead{[O II]3727} & \colhead{[O III]4363} & \colhead{H$\beta$} & \colhead{[O III]4959} & \colhead{H$\alpha$} & \colhead{[N II]6583} & \colhead{[S II]6717} & \colhead{[S II]6731}}
\startdata
O3ELG1 & $24.38\pm2.25$ & $3.22\pm1.47$ & $27.38\pm3.27$ & $47.75\pm3.83$ & $78.11\pm14.01$ & $-0.29\pm0.72$ & $-0.33\pm0.83$ & $-0.38\pm1.02$ \\
O3ELG2 & $152.53\pm20.77$ & $18.52\pm3.04$ & $141.61\pm18.88$ & $273.70\pm34.21$ & $196.50\pm32.25$ & $4.03\pm0.95$ & $3.07\pm1.02$ & $3.43\pm0.88$ \\
O3ELG3 & $6.54\pm2.93$ & $3.71\pm4.13$ & $11.13\pm5.06$ & $14.74\pm6.91$ & $31.71\pm18.25$ & $0.58\pm1.22$ & $-0.52\pm0.95$ & $-4.13\pm1.93$ \\
O3ELG4a & $39.51\pm4.80$ & $2.60\pm0.86$ & $29.23\pm4.59$ & $46.60\pm7.18$ & $82.44\pm24.32$ & $1.07\pm0.43$ & $4.40\pm0.90$ & $4.16\pm0.88$ \\
O3ELG4b & $81.01\pm25.15$ & $4.67\pm2.67$ & $27.67\pm11.22$ & $22.83\pm9.22$ & $69.64\pm40.75$ & $3.72\pm2.37$ & $7.37\pm3.85$ & $5.95\pm4.32$ \\
O3ELG5 & $34.47\pm4.96$ & $5.14\pm2.28$ & $18.53\pm3.33$ & $35.73\pm4.96$ & $68.65\pm14.43$ & $1.92\pm0.71$ & $1.43\pm1.79$ & $3.81\pm0.90$ \\
O3ELG6 & $9.87\pm2.51$ & $-1.04\pm2.24$ & $8.76\pm1.72$ & $18.87\pm2.04$ & $49.91\pm9.34$ & $1.39\pm0.73$ & $2.91\pm0.73$ & $1.26\pm0.83$ \\
O3ELG7 & $2.14\pm5.65$ & $6.43\pm2.24$ & $48.45\pm10.26$ & $106.49\pm17.72$ & $149.31\pm31.28$ & $1.71\pm0.30$ & $0.43\pm0.85$ & $3.02\pm0.66$ \\
O3ELG8 & --- & --- & --- & --- & $18.99\pm3.25$ & $2.60\pm1.14$ & $3.51\pm1.43$ & $2.17\pm1.03$ \\
O3ELG9 & $370.34\pm91.16$ & $130.25\pm43.51$ & $991.73\pm236.08$ & $1575.69\pm372.60$ & $2954.01\pm637.27$ & $16.52\pm4.68$ & $47.52\pm11.11$ & $29.47\pm9.85$ \\
O3ELG10 & $-2.80\pm1.27$ & $-1.99\pm1.98$ & $5.97\pm0.67$ & $3.29\pm0.57$ & $11.57\pm1.97$ & $1.23\pm0.61$ & $1.44\pm0.81$ & $-0.48\pm0.60$ \\
O3ELG11 & $81.99\pm21.83$ & $12.10\pm2.90$ & $77.94\pm19.76$ & $172.65\pm41.41$ & $240.15\pm61.59$ & $6.64\pm1.66$ & $12.41\pm3.01$ & $8.98\pm2.68$ \\
O3ELG12a & $48.23\pm21.30$ & $-5.38\pm1.89$ & $27.33\pm9.31$ & $32.13\pm10.60$ & $152.57\pm36.11$ & $3.67\pm2.21$ & $11.99\pm2.37$ & $12.56\pm3.82$ \\
O3ELG12b & $24.23\pm2.59$ & $4.69\pm1.54$ & $41.48\pm4.35$ & $59.74\pm5.79$ & $109.88\pm16.19$ & $0.63\pm0.55$ & $2.53\pm0.24$ & $1.25\pm0.59$ \\
O3ELG13 & $102.46\pm28.35$ & $6.79\pm2.90$ & $97.85\pm32.09$ & $185.34\pm57.38$ & $306.86\pm83.00$ & $2.61\pm1.14$ & $14.82\pm4.62$ & $8.29\pm2.98$ \\
O3ELG14 & $59.27\pm18.70$ & $3.29\pm3.03$ & $79.15\pm26.74$ & $110.81\pm36.04$ & $220.59\pm63.72$ & $3.79\pm1.00$ & $9.43\pm2.66$ & $1.30\pm0.52$ \\
O3ELG15 & $203.13\pm39.33$ & $33.25\pm12.44$ & $120.03\pm25.23$ & $226.88\pm47.13$ & $209.38\pm56.02$ & $3.31\pm1.00$ & $10.74\pm2.57$ & $7.19\pm1.74$ \\
O3ELG16 & $12.80\pm2.54$ & $0.03\pm2.15$ & $22.75\pm4.34$ & $26.29\pm3.95$ & $63.84\pm14.45$ & $5.53\pm2.84$ & $14.95\pm3.63$ & $-1.46\pm1.63$
\enddata
\tablecomments{Fluxes are calibrated to the VIRUS [O III]$\lambda 5007$ measurement except for O3ELG8 since this object does not have an LRS2-B observation. These lines are marked with an asterisk. For this reason this table does not show a LRS2 [O III]$\lambda 5007$ flux as it is the same as the VIRUS measurement in Table~\ref{samp_tbl}.  Units are $10^{-17}$~ergs~s$^{-1}$~cm$^{-2}$. *Note O3ELG4b is not included in our sample as it did not meet our selection criteria but its properties are presented since it is a companion of O3ELG4a.}
\end{deluxetable*}

%\end{turnpage}

%\begin{turnpage}
\begin{deluxetable*}{lcccccccc}[ht!]
\tablecaption{LRS2 flux measurements for secondary lines in spectra. \label{flux_sec_tbl}}
\tablehead{\colhead{ID} & \colhead{[Ne III]3869} & \colhead{H$\delta$} & \colhead{H$\gamma$} & \colhead{He I 4471} & \colhead{He II 4686} & \colhead{He I 5876} & \colhead{[O I] 6300} & \colhead{[S III]6312}}
\startdata
O3ELG1 & $2.07\pm2.07$ & $7.87\pm1.19$ & $11.73\pm2.61$ & $0.82\pm1.04$ & $4.00\pm1.03$ & $2.27\pm1.35$ & $0.12\pm0.50$ & $0.99\pm0.80$ \\
O3ELG2 & $54.39\pm8.64$ & $13.72\pm4.38$ & $63.17\pm7.72$ & $10.74\pm2.53$ & $5.88\pm1.10$ & $16.90\pm2.49$ & $2.22\pm0.92$ & $3.18\pm1.40$ \\
O3ELG3 & $-12.82\pm5.30$ & $5.92\pm3.24$ & $4.23\pm4.10$ & $3.09\pm3.59$ & $1.19\pm4.37$ & $0.32\pm1.31$ & $-0.94\pm0.92$ & $0.06\pm0.54$ \\
O3ELG4a & $10.58\pm1.82$ & $2.21\pm1.40$ & $9.79\pm3.30$ & $-0.26\pm1.85$ & $1.78\pm1.10$ & $4.05\pm0.96$ & $1.50\pm0.85$ & $1.28\pm0.76$ \\
O3ELG4b & $7.20\pm2.81$ & $3.70\pm1.56$ & $6.82\pm3.17$ & $1.34\pm1.76$ & $2.53\pm1.55$ & $2.98\pm2.16$ & $2.61\pm1.33$ & $-0.82\pm0.57$ \\
O3ELG5 & $6.24\pm3.26$ & $-0.33\pm1.84$ & $6.96\pm1.81$ & $-0.60\pm2.39$ & $1.47\pm2.81$ & $2.40\pm0.72$ & $1.06\pm0.79$ & $-0.20\pm0.70$ \\
O3ELG6 & $7.54\pm1.45$ & $6.93\pm1.10$ & $11.31\pm2.62$ & $1.71\pm1.99$ & $-1.36\pm1.28$ & $2.50\pm0.61$ & $0.49\pm0.73$ & $0.38\pm0.79$ \\
O3ELG7 & $23.21\pm3.99$ & $5.41\pm3.85$ & $26.17\pm4.11$ & $3.23\pm1.99$ & $0.11\pm1.37$ & $3.52\pm0.90$ & $2.16\pm1.19$ & $1.88\pm0.49$ \\
O3ELG8 & --- & --- & --- & --- & --- & --- & $2.92\pm1.15$ & $0.67\pm1.24$ \\
O3ELG9 & $289.31\pm60.38$ & $215.36\pm40.96$ & $461.09\pm176.12$ & $33.34\pm13.33$ & $24.92\pm6.75$ & $84.35\pm18.50$ & $10.97\pm2.25$ & $9.85\pm3.49$ \\
O3ELG10 & $2.84\pm1.45$ & $2.87\pm1.01$ & $-1.66\pm3.17$ & $-5.54\pm1.20$ & $-0.70\pm0.70$ & $-0.60\pm0.77$ & $-0.23\pm0.83$ & $1.64\pm0.64$ \\
O3ELG11 & $36.80\pm8.79$ & $9.75\pm3.05$ & $27.48\pm7.15$ & $3.99\pm1.37$ & $-1.29\pm1.05$ & $7.47\pm2.02$ & $-0.46\pm1.40$ & $-0.51\pm0.50$ \\
O3ELG12a & $5.82\pm4.41$ & $-11.49\pm5.17$ & $2.00\pm3.10$ & $-7.67\pm7.13$ & $-11.15\pm4.82$ & $-0.07\pm4.84$ & $1.52\pm1.50$ & $-1.79\pm2.36$ \\
O3ELG12b & $14.84\pm2.02$ & $10.21\pm1.94$ & $-9.11\pm2.82$ & $-0.09\pm1.55$ & $1.26\pm1.25$ & $-0.09\pm1.19$ & $1.29\pm0.62$ & $0.68\pm0.50$ \\
O3ELG13 & $40.68\pm10.73$ & $27.43\pm6.27$ & $49.45\pm17.33$ & $3.19\pm1.39$ & $1.98\pm1.35$ & $11.57\pm3.80$ & $3.48\pm0.85$ & $-0.32\pm0.48$ \\
O3ELG14 & $21.74\pm6.82$ & $18.55\pm5.16$ & $39.13\pm12.28$ & $-2.27\pm2.43$ & $0.30\pm1.08$ & $5.21\pm2.01$ & $1.09\pm0.41$ & $1.07\pm0.48$ \\
O3ELG15 & $95.31\pm21.42$ & $32.61\pm11.60$ & $101.42\pm36.17$ & $7.60\pm4.55$ & $4.91\pm3.09$ & $9.68\pm2.50$ & $1.70\pm0.47$ & $-1.07\pm0.60$ \\
O3ELG16 & $2.36\pm0.87$ & $7.86\pm1.70$ & $3.50\pm5.43$ & $2.33\pm2.52$ & $-0.80\pm3.85$ & $4.96\pm9.88$ & $-2.89\pm2.61$ & $-0.60\pm1.84$ \\
\enddata
\tablecomments{Fluxes are calibrated to the VIRUS [O III] flux shown in Table~\ref{samp_tbl} except for O3ELG8 since this object does not have an LRS2-B observation. These lines are marked with an asterisk. Units are $10^{-17}$~ergs~s$^{-1}$~cm$^{-2}$. *Note O3ELG4b is not included in our sample as it did not meet our selection criteria but its properties are presented since it is a companion of O3ELG4a.}
\end{deluxetable*}

%\end{turnpage}

\begin{figure}[ht!]
\epsscale{1.0}
\plotone{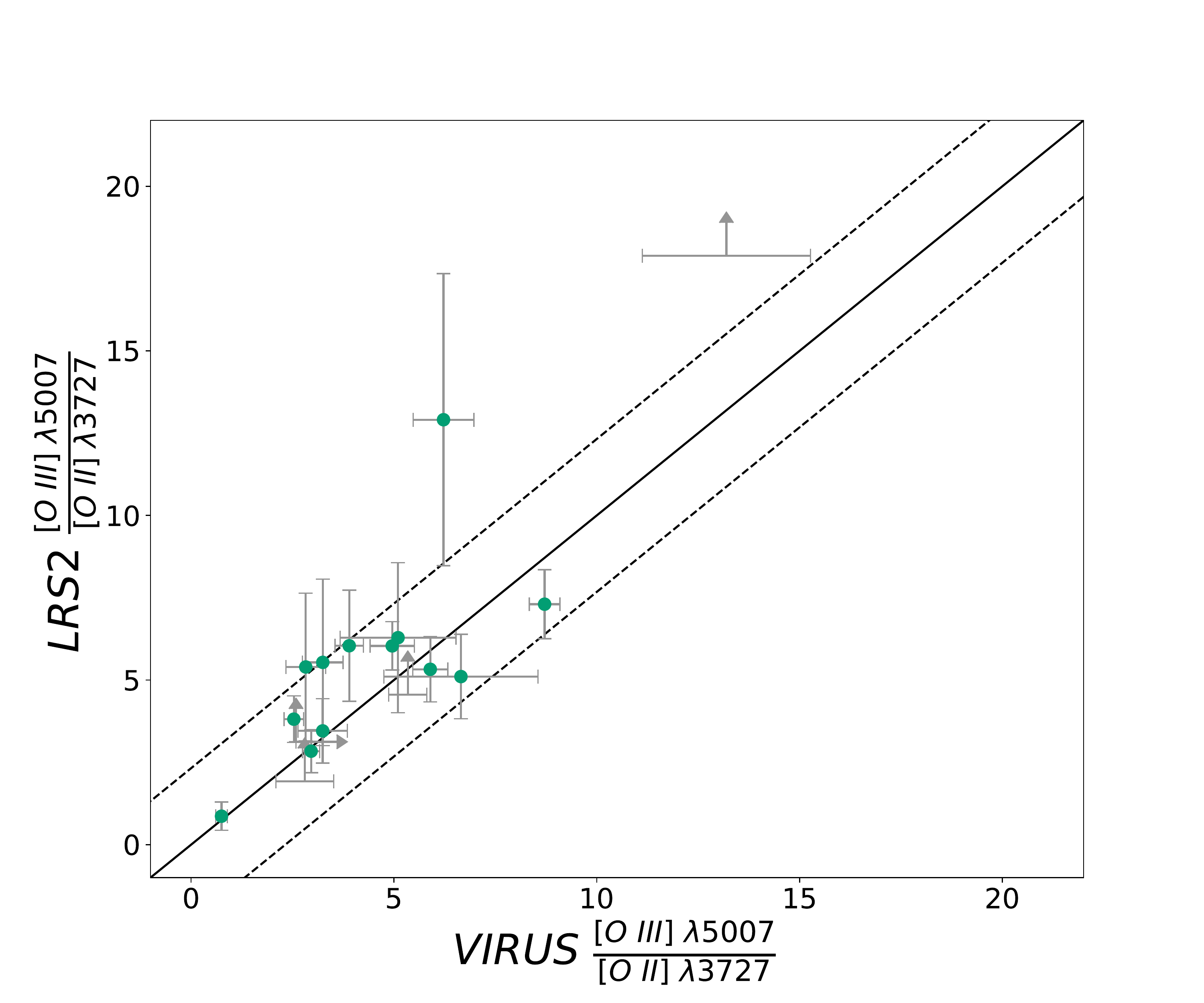}
\caption{A comparison between the \OIII $\lambda 5007$/\OII $\lambda 3727$ (O3O2) ratios measured with the VIRUS and LRS2 line fluxes. The dashed lines on either side of the 1:1 relation represent the 1$\sigma$. }.\label{fig:flux_compare}
\end{figure}

This study is designed to serve as a pilot program for building a complete sample from HETDEX\null. In the much larger future study, it will not be possible to obtain supplementary LRS2 observations for every candidate; we will be restricted to using just the VIRUS data, which covers 3500~\AA\ to 5500~\AA\ at a resolution lower than LRS2\null. We first compare the  \OIII $\lambda 5007$/\OII $\lambda 3727$ (O3O2) ratios measured with VIRUS to the corresponding LRS2 line fluxes; the results are presented in Figure~\ref{fig:flux_compare}. Ratios of galaxies with an \OII $\lambda 3727$ signal-to-noise less than three are represented as lower limits. The dotted line on either side of the 1:1 line represents the 1$\sigma$ error of these ratios, excluding the objects represented as lower limits. The one object falling furthest outside of the 1$\sigma$ boundary (O3ELG9) has high S/N noise lines and small error bars on the ratio for both instruments.  This result is likely due to difficult spectral extraction of the spatially extended object. See Figure~\ref{fig:spec_p2} in Appendix~\ref{sec:data_panels} for and \OIII map from VIRUS showing the extent of this object.  \par

\section{Deriving Galaxy Properties} 
\label{sec:prop}

We infer the metallicity, stellar mass, star formation rate, and \OIII $\lambda 5007$ equivalent width (EW) for the 17 galaxies in our sample. These properties are directly tied to their evolution and are used to compare our sample with photometrically-selected samples of similar populations to place our galaxies in context. 

\subsection{Metallicity}
\label{sec:met}

Strong line ratios (SLRs) of collisionally-excited lines to hydrogen Balmer lines are the most commonly used method for inferring gas phase metallicities, as they are relatively easy to derive observationally. Many studies have calibrated SLRs to samples with known metallicities. However, it has been well documented in the literature that large offsets in inferred metallicities are found between different SLRs when only a single SLR calibration is used (See \citet{kew08} for a review of this topic).  To address this issue, studies such as \citet{mai08} and \citet{cur17} have developed improved methods for using strong lines by calibrating several commonly used SLRs over a wide range of log(O/H)+12.  These methods allow log(O/H)+12 to be inferred from several SLRs simultaneously. \citet{mai08} calibrated their SLRs using a sample of 259 galaxies with metallicities measured with the $T_e$ direct method. This calibration focuses on the low metallicity regime:  for the direct method to work, the temperature sensitive auoral lines, such as \OIII $\lambda 4363$, must be accurately measured, and these features become extremely faint with increasing metal abundance. To calibrate SRLs at log(O/H)+12 $>$ 8.3, \citet{mai08} used a sample of 22482 galaxies with metallicities inferred from photoionization models. \citet{cur17} also built a system of calibrated SLRs, and extended the use of the more reliable $T_e$ direct abundance measurements to the higher metallicity regime by stacking over 900000 galaxies in bins based on their  \OII $\lambda 3727$/H$\beta$ and \OIII $\lambda 5007$/H$\beta$ ratios. Using a system of these calibrated SLRs together provides a more reliable method for inferring log(O/H)+12 from strong lines. \par

We model our approach after that described by \citet{gra16}. This method for inferring gas phase metallicity was also adopted in \citet{ind19}.  For each of our objects, we infer gas phase metallicities using a Bayesian approach that simultaneously fits log(O/H)+12 and E(B-V) for the system of SLRs given by \citet{mai08}. We adopted SLRs from \citet{mai08} since their models are calibrated to lower metallcities where our sources are expected to lie. Although the \citet{cur17} study is more recent and has an improved calibration in the high metallicity regime, their models are only calibrated down to log(O/H)+12 $\sim 7.6$. We use the following subset of  \citet{mai08} SLR relations:

\begin{equation}
R23\footnote{\citet{pil05, nag06}.}=\frac{{\rm [O~III]}\lambda5007+{\rm [O~III]}\lambda4959+{\rm [O~II]}\lambda3727}{{\rm H}\beta},
\end{equation}

\begin{equation}
{\rm O3O2}\footnote{\citet{kew02, nag06, bia18}.}=\frac{{\rm [O~III]}\lambda5007}{{\rm [O~II}]\lambda3727}
\end{equation}

\begin{equation}
{\rm N2}\footnote{\citet{den02, pet04, nag06}.} = \frac{{\rm [NII]} \lambda6583}{{\rm H}\alpha}
\end{equation}
\par

% Our log likelihood function is defined as the log of $\chi^2$ of the models compared to our observables in the following form:
% \begin{equation}
% {\chi}^{2} = \sum_{n=1}^{{\rm \# models}}{\frac{(x_{\rm obs} - x_{\rm mod})^{2}}{\sqrt{{\sigma^{2}}_{\rm obs} +{\sigma^{2}}_{\rm mod}}}}
% \end{equation}
% where $x_{\rm obs}$ are the observed emission line fluxes, $x_{\rm mod}$ are the model fluxes, $\sigma_{\rm obs}$ is the measurement error, and $\sigma_{\rm mod}$ represents the model flux errors given by the \citet{mai08} SRL models. 

Our log likelihood function defines the observed quantities as the observed emission line fluxes, while the models are model fluxes derived from the \citet{mai08} SRL models. This function sums over each of the \citet{mai08} SRL models used. In order to compare our measured emission line fluxes to model fluxes, we must convert the \citet{mai08} SLR relations into flux models. The \citet{mai08} models are in the form $y=f(x)$ where $y$ is a ratio of emission lines and $x$ is the metallicity. In order to express these relationships in terms of flux instead of flux ratios, an \OIII $_{\rm intrinsic}$ normalization must be defined. Since the red lines in the N2 ratio are observed with a different channel of LRS2 than the blue lines, we are not always able to calibrate the two observations well enough to reliably measure ratios between the red and the blue or measure E(B-V) from the Balmer decrement. To avoid a potentially large error in the red to blue calibration, the red lines are given their own normalization, H$\alpha_{\rm intrinsic}$, as a parameter to be fit. $E(B-V)$ is then left as another parameter in the model to be fit by our code.  The model fluxes are multiplied by a reddening factor ${10}^{-0.4~K(\lambda)~E(B-V)}$ where $K(\lambda)$ is the \citet{cal00} attenuation curve at the wavelength of the line and E(B-V) is a parameter to be marginalized over in the fit. Each model flux is then a function of the fit parameters, metallicity, $E(B-V)$ and either \OIII $_{\rm intrinsic}$ or H$\alpha_{\rm intrinsic}$. \par

Flat priors are set on metallicity that require the fit to be between a $7.0 < \log({\rm O/H})+12 < 10.0$. \OIII $_{\rm intrinsic}$ and H$\alpha_{\rm intrinsic}$ also have flat priors that bound them between 0.0 and 10$^{-11}$~ergs~cm$^{-2}$~s$^{-1}$. A Gaussian prior was used for $E(B-V)$ with a mean of 0.112 and a standard deviation of 0.086. These values are based on the distribution of E(B-V) values that were measured in the HII regions of a sample of typical dwarf galaxies from \citet{ber12}. Moreover, since the reddening could be close to zero, we must allow for the values to go slightly negative (within the uncertainty) so we do not systematically overestimate the reddening of the sample, and thus bias our results  towards lower metallicities. As a test, we ran the code with Gaussian prior on $E(B-V)$ set to the mean (0.295) and standard deviation (0.165) of the SDSS sample of typical star-forming galaxies.  We then compared these results to those run $E(B-V)$ fixed at 0.0.  In almost all cases, the metallicities returned by the two assumptions agreed with each other within the errors. \par

\begin{figure}[ht!]
\epsscale{1.0}
\plotone{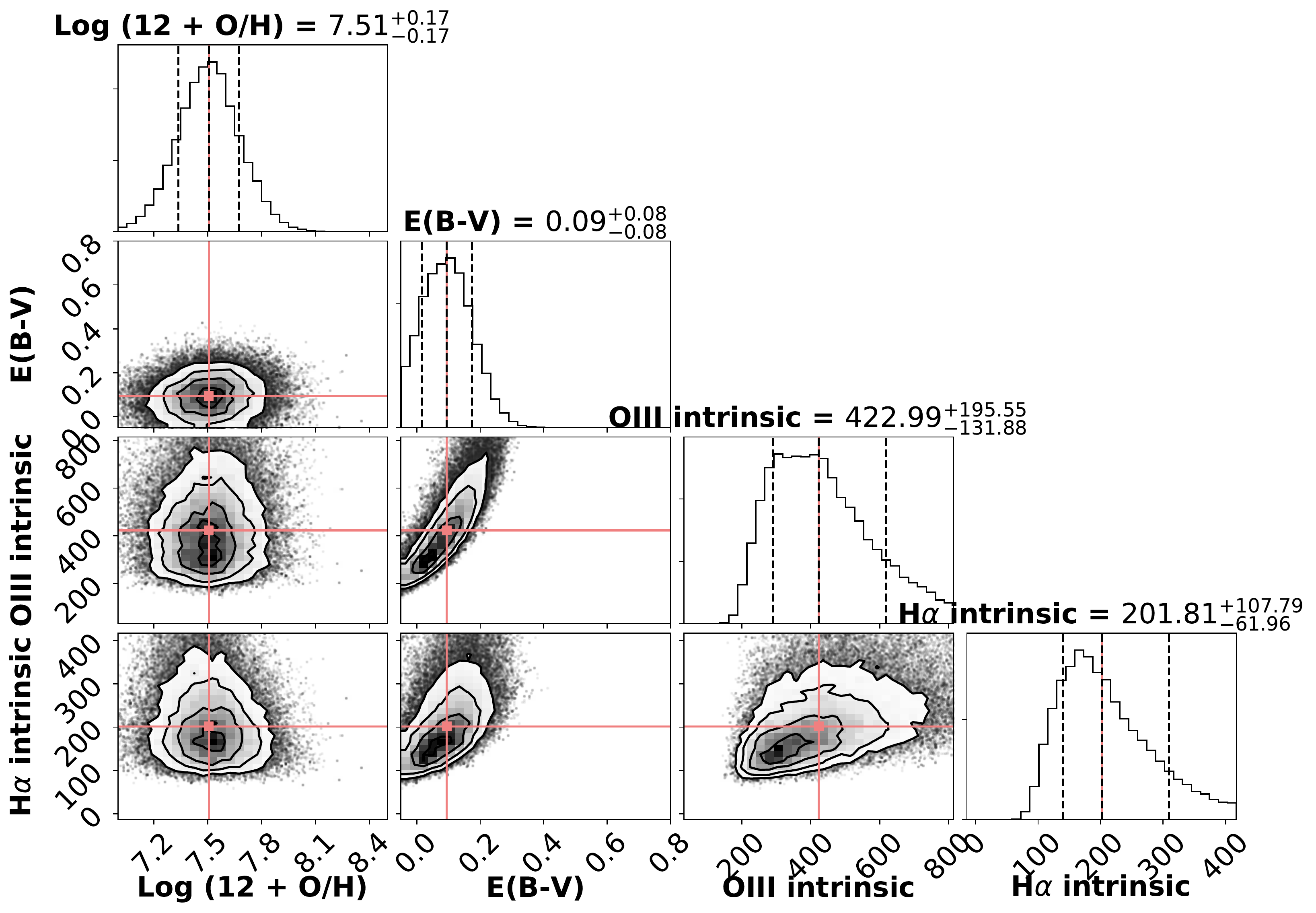}
\caption{An example output from the metallicity fitting code showing the posterior of the 4 parameters fit by the code.}.\label{fig:cornerplot}
\end{figure}

\begin{figure*}[ht!]
\epsscale{1.0}
\plotone{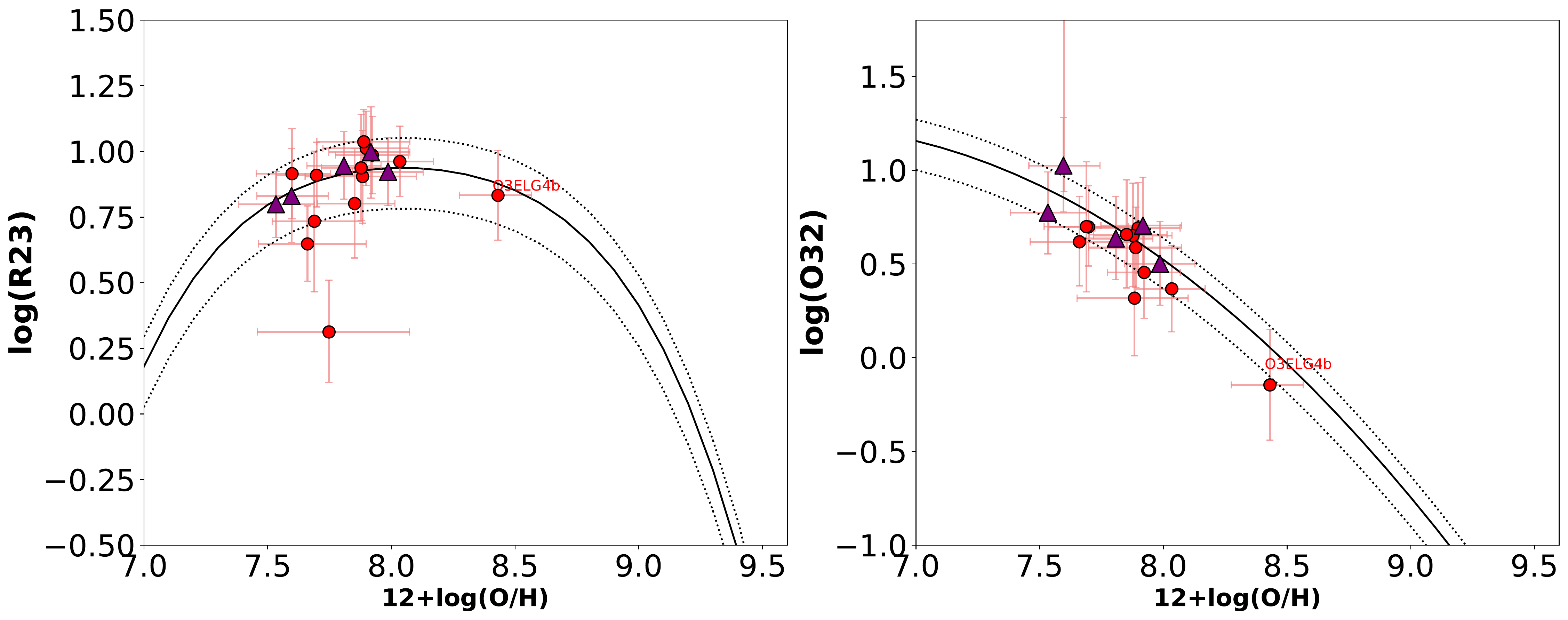}
\caption{The \citet{mai08} relations (solid black line) with 1$\sigma$ error bars (dashed lines) for the strong line ratios R23 and O3O2. The points show the flux ratios (corrected for reddening based on the $E(B-V$) fit from our Bayesian code) and the inferred galaxy metallicities.  Objects shown with a purple triangle have an  \OIII $\lambda 4363$ measurement sufficiently robust to infer log(O/H)+12 from the $T_e$ direct method. Note O3ELG4b is no included in our sample as it did not meet our selection criteria but it is presented here (labeled O3ELG4b) since it is a companion of O3ELG4a. It is clear that it is a much higher metallicity source compared to our sample. }.\label{fig:maiolino}
\end{figure*}

The solution was taken to be the 50th percentile value of the posterior for the parameters with the upper and lower 1$\sigma$ errors defined to be the 84th-50th percentile and 50th-16th percentiles, respectively. 
%An example posterior solution is shown in Figure~\ref{cornerplot}.  
Figure~\ref{fig:maiolino} compares the R23 and O3O2 \citet{mai08} SLR models in black with our output metallicity values overplotted in red. The $x$-axis represents the log(O/H)+12 values fit from our code. The $y$-axis gives the SLRs calculated from our observed fluxes corrected for reddening using the $E(B-V)$ inferred from our code. \par

For five of the objects,  \OIII $\lambda 4363$ is detected in the LRS2 data above 5$\sigma$. For these galaxies we additionally infer a metallicity using the electron temperature ($T_e$) direct method. This technique, which is based on physics rather than an empirical relation, should be the more reliable derivation of oxygen abundance.  In most cases,  \OIII $\lambda 4363$ is too faint for an accurate measurement of electron temperature, but for these five galaxies, we could measure $T_e$ using the  \OIII $\lambda 4363$/\OIII $\lambda 5007$ ratio.  We did this using \texttt{python} package \texttt{PyNeb},  which provides nebular diagnostic tools using the equations based on \citet{ost89}. Since electron temperature depends on electron density,  \texttt{PyNeb} was used to estimate the electron density using the  \SII $\lambda\lambda 6717,6731$ ratio. The five objects have measured \SII ~electron densities of approximately $10^2$~cm$^{-3}$ or smaller, making the contribution to the electron temperature measurement negligible. The $12 + \log {\rm O}^+/{\rm H}$ and $12 + \log {\rm O^{+2} / {\rm H}}$ were calculated with the equations in \citet{izo06} based on photoionization models of \ion{H}{2} galaxies, and the total oxygen abundance was calculated from:
\begin{equation}
{\rm \frac{O}{H}} = {\rm \frac{O^+}{H^+}} + {\rm \frac{{O}^{2+}}{H^+}}
\end{equation}
\par

We do not include a term to account for higher ionization species of Oxygen in either method. \citet{izo06} notes that ${\rm O}^{3+}/\rm O \gtrsim 1\%$ only when ${\rm O}^{+}/({\rm O}^{+}+{\rm O}^{2+}) \lesssim 0.1$ in the highest-excitation HII regions. We estimate the ${\rm O}^{+}/({\rm O}^{+}+{\rm O}^{2+})$ ratio for all galaxies with a \OIII $\lambda 4363$ S/N $>$ 3.0 and find that all but two of the galaxies have ratios greater than 0.1 in the regime that the higher ionization species is negligible. We also looked at the S/N of the He II $\lambda$ 4686 in all of our galaxies since it has a similar ionization potential as triply ionized oxygen. Only two systems had He II $\lambda$ 4686 S/N $>$ 5.0. Only O3ELG9 has an ${\rm O}^{+}/({\rm O}^{+}+{\rm O}^{2+})$ ratio less than 0.1 and a significant He II $\lambda$ 4686 detection. However, we still choose to not add an additional term to account for the higher ionization species since O3ELG9 is the only system we believe has a non negligible contribution from ${\rm O}^{3+}$. However, we believe this contribution is small and within the errors of our metallicity values. \par

\begin{figure}[ht!]
\epsscale{1.0}
\plotone{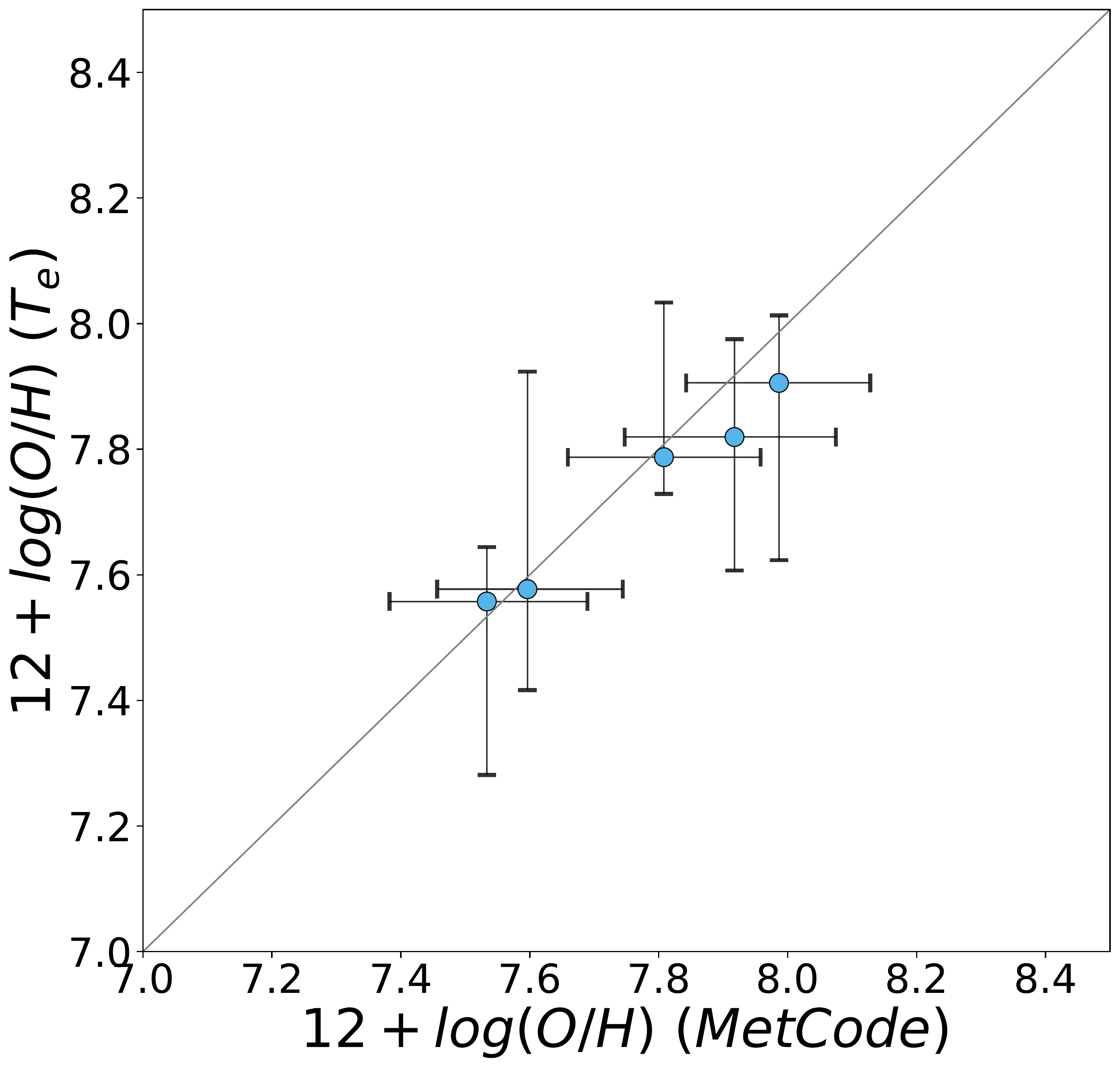}
\caption{log(O/H)+12 inferred from the \citet{mai08} strong-line indicators using our Bayesian scheme compared to metallicities we measured directly using electron temperatures measured from  \OIII $\lambda 4363$ lines with S/N$>$5.}.\label{fig:metcompare}
\end{figure}

These five $T_e$ metallicities were compared to the metallicities inferred from the SLRs in Figure~\ref{fig:metcompare}. The log(O/H)+12 values inferred from both the metallicity fitting code and the log(O/H)+12 inferred from the $T_e$ direct method are given in Table~\ref{prop_tbl}.

\subsection{Stellar Mass}
\label{sec:mass}

The stellar masses of these systems were estimated via spectral energy distribution (SED) fitting using \texttt{MCSED} \citep{bow20}. This SED fitting software employs a stellar library generated by the Flexible Stellar Population Synthesis (\texttt{FSPS}) code \citep{fsps-1, fsps-2} using PADOVA isochrones \citep{bertelli+94, girardi+00, marigo+08}, the prescription for nebular line and continuum emission given by the grid of \texttt{cloudy} models \citep{CLOUDY, CLOUDY13} generated by \citet{byler+17}, and a \citet{cha03} initial mass function. We adopted a ten-parameter model, with the variables being stellar metallicity (ranging from $\sim 1-150\%~Z_{\odot}$), a seven age-bin star formation history (using a constant star formation rate within each stellar age bin between [0.001, 0.02, 0.1, 0.3, 1.0, 3.2, 6.3, 13.2 Gyr]), and a three-parameter dust attenuation law \citep{nol09}. The ionization parameter was fixed at a value of $\log U = -2.5$; this choice had no systematic effect on the stellar mass estimates. Changing the adopted fitting assumptions (especially the star formation rate history) can systematically affect the stellar masses at the level of $\sim 0.5$~dex.

Both the photometry described in Section~\ref{sec:photo} and the emission-line fluxes from the VIRUS observations were used to constrain the SED fits by including additional $\chi^2$ terms for each flux measurement in the SED model likelihood function.  These emission lines contain information about (primarily) the systems' current star formation rates and metallicities, so using the observed line fluxes to constrain the fits narrows the posterior distributions on these parameters, yielding tighter constraints on the other model variables. The stellar mass estimates are systematically lower by a factor of $\lesssim 0.5$~dex when the emission lines are used to constrain the fits (compared to when only the photometry is used), since the H$\beta$ fluxes exclude solutions with low recent SFRs (shifting the balance in stellar populations from more numerous, older stars to fewer, younger stars).

Our systems have intense, recent star formation for low-redshift systems, as evidenced by their H$\alpha$ flux (see Figure~\ref{fig:sfr}), outshining any older stellar populations.  With the stellar continuum and nebular emission light dominated by massive stars ($>5 {M}_{\odot}$ masses), our assumptions in SED fitting play a large role in the best-fit stellar masses as well as the quoted uncertainties. Our uncertainties are determined from the posterior distributions of the fit parameters and represent the statistical errors on the fits. Since massive stars are relatively rare in production compared to their low-mass counterparts, our stellar mass estimates and errors are especially sensitive to the initial mass function, the star formation history parameterization, and the star formation history prior distributions.  In addition to the statistical errors output from \texttt{MCSED} there are known systematic errors associated with SED fitting assumptions which are difficult to quantify (see \citet{con13} review article for a comprehensive discussion on this topic). When comparing these mass estimates to other studies with different SED fitting techniques and assumptions, systematic shifts should be expected and can be as high as $\sim 0.5$~dex.

%The values for the stellar mass for our objects are taken to the be the 50th percentile of the posterior for the mass fit that is output form \ttext{MCSED}. 
The results from SED fitting demonstrate that our sample has generally low stellar masses with all galaxies having a stellar mass below $10^9~M_\odot$. This result is not unexpected considering their low metallicity. Even considering the systematic mass offsets from differing SED fitting techniques, it can generally be concluded that the masses fall in a regime occupied by dwarf galaxies. For this pilot study we were searching for the most obvious low-metallicity candidates, so we chose selection criteria requiring high S/N detections. Due to this approach we expect there to be many more, lower-mass sources in the HETDEX catalog. \par

The best fit SEDs for each source can be found in Figure~\ref{fig:sed_fits} in Appendix~\ref{sec:sed_panels}. Values for the masses can be found in Table~\ref{prop_tbl}. For O3ELG10 we were unable to assign a reliable SED fit as it only had three photometric bands to fit and flux densities in these bands are on the edge of the detection limits for the survey. Because of these limitations its SED fit properties should only be considered as order of magnitude estimates for O3ELG10.  

\subsection{Reddening}
\label{sec:red}

With optical spectroscopy we would ideally get a measure of the reddening using the Balmer decrement. However, as stated in Section~\ref{sec:met}, the 
H$\beta$ and H$\alpha$ lines are not observed simultaneously; they required different setups with LRS2-B and LRS2-R, sometimes nights apart.  As a result, creating a robust calibration between the two units is challenging, and the Balmer decrement cannot be reliably measured for all but three of the lowest redshift objects in our sample were both lines are measured in the LRS2-B channels. \par

Our metallicity code marginalizes over $E(B-V)$ providing an estimate of the reddening; however, the resulting $E(B-V)$ values are heavily dependent on the adopted prior. Note, even though we found the $E(B-V)$ values depend on the prior, the metallicity values inferred did to not change outside the metallicity error bar given the different E(B-V) priors tested.  Alternatively, our SED fits infer a value of stellar reddening, and we can translate this stellar $E(B-V)$ to a nebular $E(B-V)$ using ${E(B-V)}_{\rm stellar} = 0.44~{E(B-V)}_{\rm nebular}$ from \citet{cal01}. Despite the known systematics of this correction \citep{bat16} and with SED fitting in general, we still believe that the stellar-based $E(B-V)$ inferred from \texttt{MCSED} produces our best estimate for reddening, so these are the values we adopt to de-redden the H$\alpha$ fluxes. The $E(B-V)$ values from \texttt{MCSED} (scaled to their nebular values) are presented in Table~\ref{prop_tbl}. To test the robustness of the \texttt{MCSED} derived E(B-V) values we compared them to E(B-V) values derived with the Balmer decrements for the three objects at sufficiently low redshift and find they are consistent.

\subsection{Star Formation Rates}
\label{sec:sfr}

We estimate an instantaneous SFR from the LRS2 H$\alpha$ line flux corrected for reddening using the $E(B-V)$ from \texttt{MCSED} (with the 0.44 correction) as discussed in Section~\ref{sec:red} and an attenuation curve from \citet{cal00}. \par

To estimate the SFR of our galaxies, we adopt the SFR-H$\alpha$ luminosity relation from \citet{ken98,ken09,ken12}:
\begin{equation}
    \textrm{SFR}_{\rm{H\alpha}}(M_{\odot}/{\rm yr}) = \frac{L({\rm H\alpha})}{{\eta}_{H\alpha}}
\end{equation}
where ${\eta}_{H\alpha}$ is the conversion factor. The quantity ${\eta}_{H\alpha}$ is often assumed to be a constant, with ${10}^{41.27}$~ergs~s$^{-1}~M_{\odot}^{-1}~{\rm yr}^{-1}$ \citep{ken12}, although \citet{bri04} found ${\eta}_{H\alpha}$ can vary up to $\sim 0.4$~dex across a wide range of stellar masses. Since our sample falls at the low end of the mass function (far from the typical stellar masses of galaxies upon which this value was inferred), we choose to adopt a parameterization that lies between the constant ${\eta}_{H\alpha}$ value and the stellar masses inferred by \citet{dua17} using the likelihood distributions on ${\eta}_{H\alpha}$ over different ranges in stellar mass  \citep{bri04}:
\begin{equation}
\log({\eta}_{H\alpha}(x)) = -0.011~{x^2}~+~0.124~x+41.107
\end{equation}
where $x=\log(M/M_{\odot})$. 

A derivation of the star formation rate (SFR) over the past $\sim 100$~Myr is also inferred via \texttt{MCSED}. The quantity relies heavily on the galaxies' GALEX photometry and the H$\beta$ line fluxes.  Just as for the stellar mass, we quote the 50th percentile of the posterior of this model parameter. \par

\texttt{MCSED} does not incorporate the H$\alpha$ fluxes in its calculation, since we can not reliably tie the LRS2-R lines to the LRS2-B or VIRUS lines.  However, the estimates do use the H$\beta$ line flux as this line falls within the VIRUS spectral range. As mentioned in Section~\ref{sec:mass}, the inclusion of  H$\beta$ yields much tighter constraints on the most recent SFR than fits with photometry alone. In particular, the H$\beta$ flux imposes a lower limit on the recent SFR, shifting the estimates to slightly higher values (and lowering the stellar mass estimates by $\lesssim 0.5$~dex). Although the measured H$\alpha$ line flux does not individually contribute to the $\chi^2$ fit, it is included in the flux density of the photometric filter in which it lies. \par

%The SFRs from \ttext{MCSED} and the H$\alpha$ line flux are compared in Figure~\ref{fig:sfr_compare} showing the SFR values between the two measures almost always agree within the errors. 

The time average SFR over $\sim 100$~Myr derived from \texttt{MCSED} and the instantaneous SFR derived from H$\alpha$ line flux are listed in Table~\ref{prop_tbl}. We did not have sufficiently accurate photometery to determine reliable SED fits for O3ELG10. Because of this limitation, O3ELG10's SFRs are very uncertain since it relies solely on the $E(B-V)$ inferred from \texttt{MCSED}. 

% \begin{figure}[ht!]
% \epsscale{1.0}
% \plotone{ha_sed_sfr_compare.pdf}
% \caption{A comparison of the the SFR averaged over the past 100~Myr inferred from SED fitting versus the SFR derived from H$\alpha$. }.\label{fig:sfr_compare}
% \end{figure}

\subsection{Equivalent Width}
\label{sec:EW}
\begin{figure}[ht!]
\epsscale{1.0}
\plotone{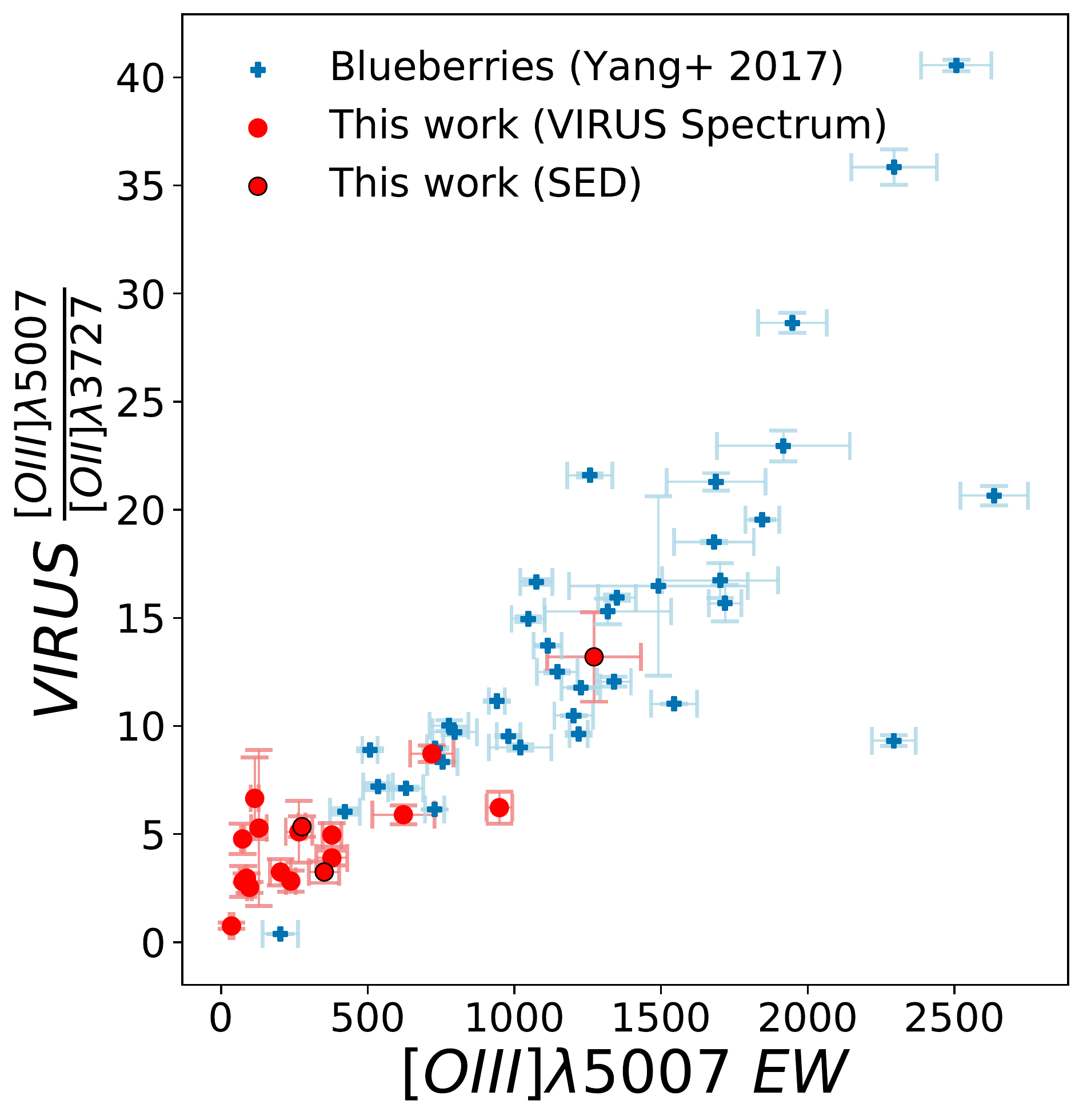}
\caption{The \OIII $\lambda 5007$ EW versus the O3/O2 ratio measured from the VIRUS spectra for our sources in the red circles. The red circles outlined in black represent objects with continuum levels not significantly detected in their VIRUS spectra. For these objects EWs are estimated from continuum levels obtained from the SED fit. For comparison the sample of blueberry galaxies, known for their extreme EWs, from \citet{yan17} is shown with the blue triangles.} \label{fig:o3_ew}
\end{figure}

We determine Equivalent widths (EWs) of \OIII $\lambda 5007$ for our sources. We measure the continuum from the VIRUS spectra at $\pm$50$\AA$ on either side of 5107.8$\AA$. For all of our objects we also estimate the continuum level from a combination of the stellar and nebular components of the SED fit. We present EWs measured from the VIRUS spectra for all sources where the continuum levels were significantly detected. For three of our galaxies where we could not measure continuum levels above a S/N$>$3 from the spectra we present EWs estimated by the SED method. We compare the EWs measured directly from the spectra themselves and found good agreement with the SED modeling method for all galaxies where EWs could be estimated from both methods.  Figure~\ref{fig:o3_ew} shows the O3O2 ratio for our sources versus the \OIII $\lambda 5007$ EW shown in red circles. The red circles that are outlined in black represent the three objects where the EW is estimated from the SED method. The same values for the blueberries from \citet{yan17} are shown in blue triangles. Our sources, for the most part,  have high EWs, however, few are as extreme as those of the blueberry sample. This is not surprising as the selection of blueberries requires them to have high EW \OIII~ in order to be found in the SDSS broad band photometry. Since our sample was selected spectroscopically we are able to find low mass sources with more moderate EW values. \OIII $\lambda 5007$ EW values for each object in our sample are presented in Table~\ref{prop_tbl}. Objects with EW values are estimated from the SED method are marked with an asterisk. 

\section{Analysis} 
\label{sec:analysis}

We analyse the relation of the basic properties of our sample, log(O/H)+12 and SFR, with stellar mass to place our sample into context with comparison populations. \par 

\subsection{Mass vs. Metallicity}
\label{sec:mass_met}

\begin{figure*}[ht!]
\epsscale{1.0}
\plotone{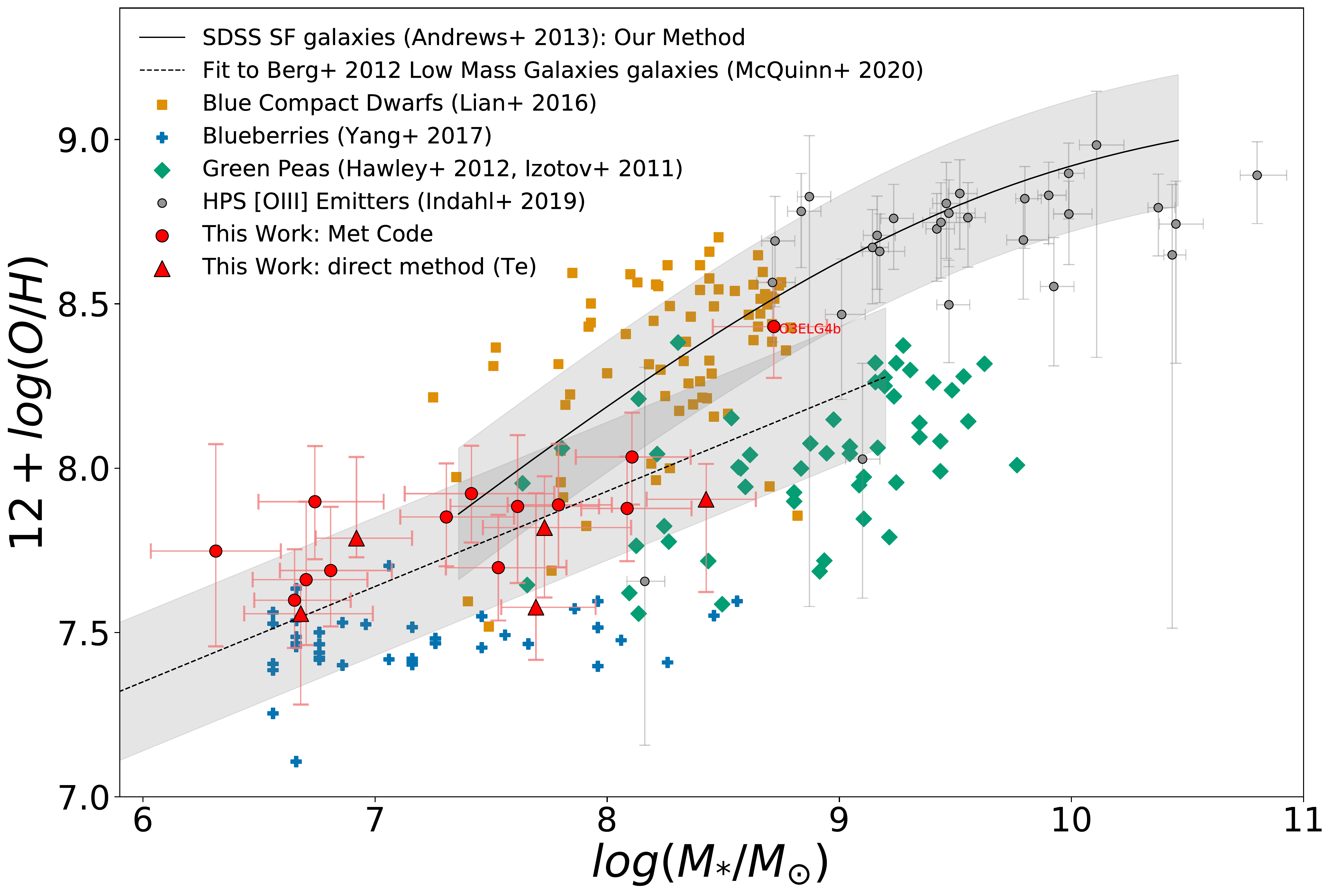}
\caption{The mass-metallicity relation for our sample of 17 low metallicity galaxies.  Solid red circles show systems with metallicities based on strong-line relations, while the red crosses denote galaxies with abundances derived directly using the object's electron temperature. Several comparison populations of photometrically-selected galaxies are included in the figure: green pea galaxies (green diamonds), blueberry galaxies (blue crosses), and blue compact dwarfs (orange squares).  The mass-metallicity relation derived from SDSS typical star-forming galaxies \citep{and13} is indicated by the black line. The shade region around this relation represents the 1$\sigma$ scatter. To extend this relation to lower mass the dotted grey line represents a fit to a sample of typical low mass dwarf galaxies from \citet{ber12}. The grey shading is the 1$\sigma$ error on the fit to the line. The \OIII $\lambda 5007$ galaxies from the HETDEX Pilot Survey \citep{ind19} are represented as grey circles. All six comparison populations have log(O/H)+12 derived with our Bayesian strong line code. Note O3ELG4b is not included in our sample as it did not meet our selection criteria but it is presented here (labeled O3ELG4b) since it is a companion of O3ELG4a. It is clear that this source is consistent with a typical field galaxy.}\label{fig:massmet}
\end{figure*}

Figure~\ref{fig:massmet} displays the mass-metallicity relation for our sample.  Our  \OIII $\lambda 5007$ selected galaxies are represented by the red circles (for systems with metallicities based on the strong line relations) and triangles (for metallicities derived directly using electron temperatures). O3LEG4b is also shown with a red circles, but labeled. This object is not included in our sample as it did not meet our selection criteria but it is presented as it is an interesting companion of O3ELG4a.\par

The solid black line represents the mass-metallicity relation derived from the $\sim 200,000$ typical star-forming galaxies detected in SDSS DR7 \citep{and13}. Comparison populations found from photometric selection such as blue compact dwarfs \citep{lia16}, green peas \citep{haw12,izo11}, and blueberries \citep{yan17} are shown with yellow squares, green diamonds, and blue crosses, respectively. To extend the mass-metallicity relation down to lower masses we fit a line to the \citet{ber12} typical star-forming low-mass galaxies, indicated by the gray dotted line. The shaded region represents the 1$\sigma$ error of the line fit. To properly compare all the populations, we have re-derived the metallicities for the green peas, blueberries, and blue compact dwarfs using our Bayesian metallicity code and fluxes from the literature. Metallicities were also re-derived using our code for each mass bin for the \citet{and13} SDSS mass-metallicity relation.  All metallicities for the low-mass typical star-forming galaxies were derived with the $T_e$ direct method, and since the \citet{mai08} relations for low-mass galaxies were calibrated using this method, we expect these values are also on the same system. The spectroscopically-selected, complete sample of \OII $\lambda 3727$ and \OIII $\lambda 5007$ galaxies found in the HETDEX Pilot Survey (HPS) from \citet{ind19} is also plotted in the grey circles. \par

The majority of this sample is fairly typical of field galaxies in terms of its mass-metallicity, however, two sources were found to fall in a low metallicity regime for their mass. Finding two sources with very low metallicity for their mass in such a small area (163 ${\rm arcminutes}^{2}$) motivated this search for low metallicity galaxies in the larger HETDEX survey as this may be evidence that spectroscopy without pre-selection of targets was successful at finding faint, low metallicity sources. \par 

The masses for the comparison populations were inferred from different methods (excluding the HPS galaxies).  While we cannot fully correct for all the systematics associated with the different assumptions used by the various SED fitting codes, we did correct each population to a common \citet{cha03} initial mass function (IMF)\footnote{To correct from \citet{kro01} to \citet{cha03} we subtract 0.04 dex \citep{muz13}. To correct from \cite{sal55} to \citet{cha03}, we multiply the stellar mass (in linear space) by 0.61 \citep{fsps-1}.}.\par

As discussed in Section~\ref{sec:mass}, there is considerable uncertainty in comparison of the stellar mass values since each sample uses a different method to find the mass. However, comparing the absolute g-band magnitudes of our objects (shown in the first column of Table~\ref{phot_tbl}) with that of the \citet{yan17} blueberry galaxies we see that our sample tends to have similar or lower g-band absolute magnitudes implying similar or lower mass. This is consistent with the comparison seen in the figure. \par   

While a couple of our sources fall close to the somewhat more extreme blueberry regime of  mass-metallicity space, these objects as a sample are consistent with typical field dwarf galaxies in terms of their mass and metallicity. Considering we are specifically searching for low metallicity sources it is not surprising these objects end up being low mass dwarfs. With spectroscopic selection this study is also not limited to finding only very high EW \OIII emitters like in photometric searches for blueberries and green peas, so it make sense that in the small area surveyed in HDR1 we primarily find typical low mass dwarfs with moderately high EWs compared to the very extreme blueberry galaxies.   \par

\subsection{Mass vs. Star Formation Rate}
\label{sec:mass_sfr}

\begin{figure*}[ht!]
\epsscale{1.0}
\plotone{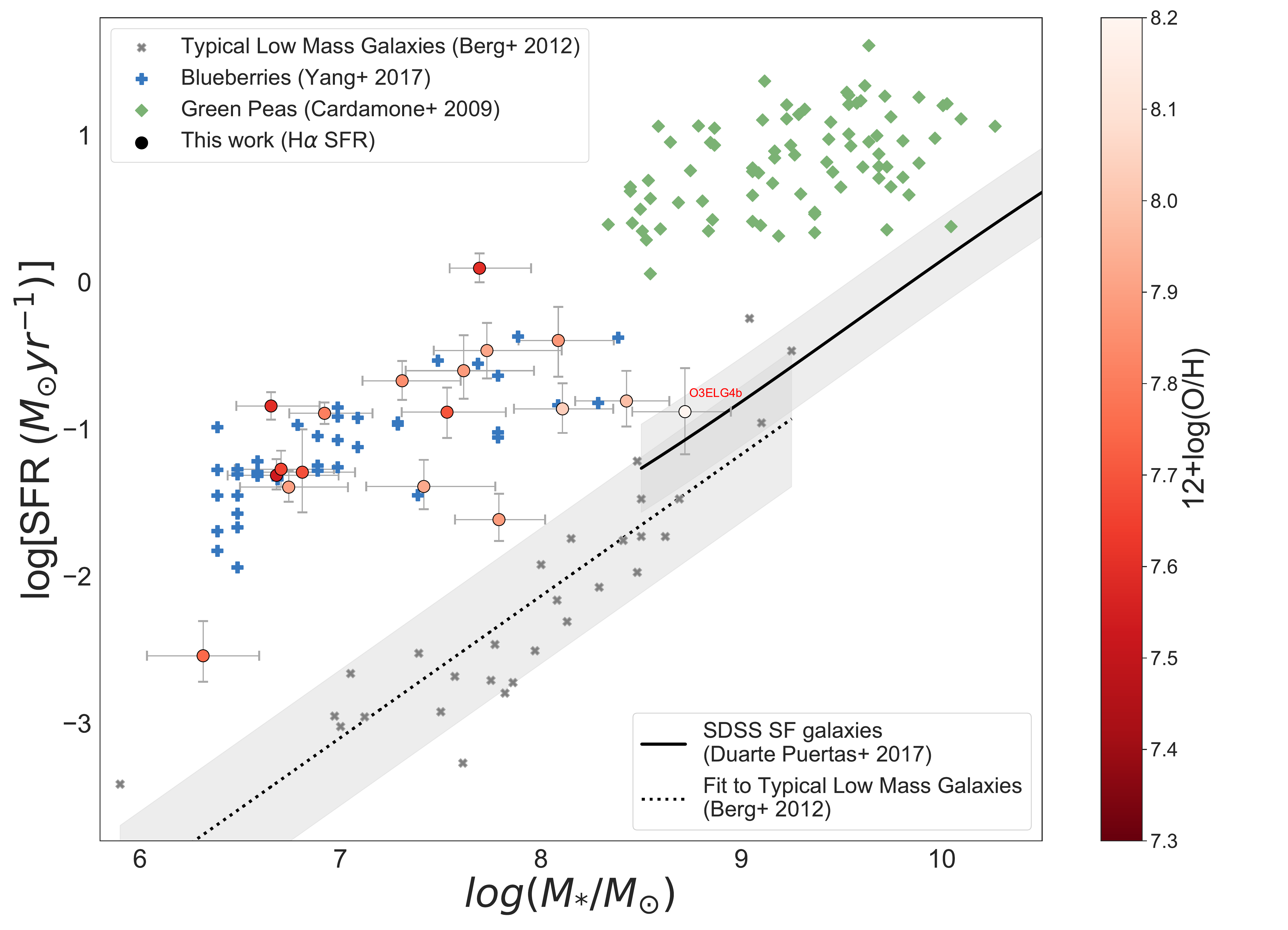}
\caption{The instantaneous H$\alpha$ star formation rates of \OIII\ objects versus their stellar mass. Our objects are represented by color-coded (as shown in the colorbar) circles representing their metallicities. Two comparison photometric-selected populations, the green peas and blueberries, are represented in green diamonds and blue crosses, respectively.Green pea and blueberry SFRs are calculated from their literature values in the same way as our objects. The solid black line represents the star-forming sequence for normal SDSS galaxies \citep{dua17}; the dotted black line represents the star-forming sequence for a sample of low mass dwarf galaxies from \citet{ber12}. The grey shaded regions represent the 1$\sigma$ error on the relation for their respective line.  All comparison populations have their masses corrected to represent the \citet{cha03} initial mass function. Note O3ELG4b is no included in our sample as it did not meet our selection criteria but it is presented here (labeled O3ELG4b) since it is a companion of O3ELG4a. It is clear that this source is consistent with a typical field galaxy.}\label{fig:sfr}
\end{figure*}

Figure~\ref{fig:sfr} shows the instantaneous (H$\alpha$) SFR versus stellar mass relation for our sample with circle markers scaled by the metallicity value. Several photometrically-selected populations are also overplotted for comparison. The solid line from \citet{dua17} represents the star-forming sequence of SDSS galaxies. Like most star-forming sequences in the literature, this relation does not reach masses below $\log(M/M_{\odot})$ of 8.5 and is based solely on continuum-selected galaxies. For a comparison at lower masses, the dotted line represents the star-forming sequence derived from a sample of low mass dwarf galaxies \citep{ber12}. Blueberries \citep{yan17} and green peas \citep{car09} samples are added as blue crosses and green diamonds, respectively. \par

There are well-studied discrepancies when using different SFR indicators and methods \citep{ken12}. In order to properly compare our sample with the other populations, we took data from the literature and re-derived the galaxies' H$\alpha$ SFRs with the technique discussed in Section~\ref{sec:sfr}. The H$\alpha$ fluxes published in \citet{ber12} are for specific HII regions within the galaxies. To avoid underestimating the SFRs we used H$\alpha$ luminosities from \citet{ken08} for the \citet{ber12} sources. A weighted average of the E(B-V) of the galaxies HII regions and the distances to the sources published in \citet{ber12} were used to correct the \citet{ken08} luminosities for reddening. The masses could not be re-derived with the same technique for all samples, as the assumptions which are used in these estimates are quite varied.  We did, however, correct all the values to a common IMF defined by \citet{cha03}. \par

From Figure~\ref{fig:sfr} it is apparent that blueberries, green peas, and our own \OIII $\lambda 5007$ galaxies all have high SFRs for their mass. Our objects appear to have similar specific star formation rates (sSFRs) (SFR divided by stellar mass) to the sample of blueberry galaxies even though they differ in terms of other properties. Despite the large errors in our \texttt{MCSED} derived E(B-V) values used to de-redden the H$\alpha$ flux we find the errors on our SFRs are small enough that our galaxies still consistently have much higher sSFRs than the typical low mass galaxies in the \citet{ber12} sample. Even if we are overestimating the E(B-V) values, our sample will still have generally higher sSFRs than typical. It is also important to note that \citet{yan17} do not correct their fluxes or SFRs for reddening for the blueberry sample and just assume these objects likely have low E(B-V) values anyway. This means the SFRs for the blueberry sample shown in Figure~\ref{fig:sfr} are likely a bit higher than those shown. Though even assuming the blueberry sample has E(B-V) values that fall on the high end of those for our sample, their SFRs would still be generally consistent with our sample. Galaxies found in HETDEX are selected purely from their strong emission lines. Within a given volume, galaxies with the strongest emission lines will likely have some of the strongest SFRs.  As a result, it is not unexpected that a volume-limited spectroscopic survey without pre-selection would preferentially find galaxies with higher than typical current SFRs. \par

%This result is not surprising considering our galaxies are all selected to have strong Oxygen emission lines which has been shown to correlate with SFR \citep{mou06}. \par

\begin{deluxetable*}{cccccccccc}[ht!]
\rotate
\tablecaption{Properties of the [O III] emitting galaxies in the sample. \label{prop_tbl}}
\tablehead{\colhead{ID} & \colhead{[OIII]/[OII]}  & \colhead{log(O/H)+12} & \colhead{log(O/H)+12}  & \colhead{SFR($M_\odot$yr$^{-1})$} & \colhead{SFR($M_\odot$yr$^{-1})$} & \colhead{Log(M/$M_\odot$)} & \colhead{sSFR(yr$^{-1})$} & \colhead{E(B-V)} & \colhead{([OIII]$\lambda$5007) EW} \\ 
\colhead{} & \colhead{(VIRUS)} & 
\colhead{(Met Code)} & \colhead{($T_e$)} & \colhead{(H$\alpha$)} & 
\colhead{(MCSED (100Myr))}  & \colhead{(MCSED)} & \colhead{(H$\alpha$)} & \colhead{(MCSED)} & \colhead{(VIRUS/MCSED$^{*}$)}}
\startdata
O3ELG1 & $4.96\pm0.55$ & ${7.70}_{0.16}^{0.16}$ & --- & ${0.13}_{0.05}^{0.05}$ & ${0.14}_{0.02}^{0.03}$ & ${7.53}_{0.29}^{0.23}$ & ${-8.41}_{0.29}^{0.34}$ & ${0.15}_{0.11}^{0.12}$ & $377.72\pm32.08$ \\
O3ELG2 & $5.90\pm0.43$ & ${7.81}_{0.15}^{0.15}$ & ${7.787}_{0.06}^{0.25}$ & ${0.13}_{0.02}^{0.02}$ & ${0.04}_{0.00}^{0.00}$ & ${6.92}_{0.24}^{0.18}$ & ${-7.81}_{0.19}^{0.25}$ & ${0.03}_{0.01}^{0.01}$ & $621.51\pm106.09$ \\
O3ELG3 & $5.34\pm0.47$ & ${7.69}_{0.17}^{0.19}$ & --- & ${0.05}_{0.03}^{0.03}$ & ${0.02}_{0.01}^{0.01}$ & ${6.81}_{0.26}^{0.22}$ & ${-8.10}_{0.35}^{0.39}$ & ${0.13}_{0.11}^{0.08}$ & $275.83\pm12.51^{*}$ \\
O3ELG4a & $2.53\pm0.24$ & ${7.99}_{0.14}^{0.14}$ & ${7.906}_{0.28}^{0.11}$ & ${0.16}_{0.06}^{0.07}$ & ${0.37}_{0.28}^{0.44}$ & ${8.43}_{0.21}^{0.26}$ & ${-9.23}_{0.31}^{0.30}$ & ${0.14}_{0.12}^{0.09}$ & $96.94\pm7.63$ \\
O3ELG4b & $0.75\pm0.15$ & ${8.43}_{0.16}^{0.14}$ & --- & ${0.13}_{0.09}^{0.09}$ & ${0.63}_{0.37}^{0.60}$ & ${8.72}_{0.23}^{0.26}$ & ${-9.60}_{0.39}^{0.37}$ & ${0.14}_{0.11}^{0.10}$ & $35.57\pm6.38$ \\
O3ELG5 & $2.96\pm0.21$ & ${8.03}_{0.14}^{0.14}$ & --- & ${0.14}_{0.05}^{0.06}$ & ${0.34}_{0.09}^{0.16}$ & ${8.11}_{0.25}^{0.24}$ & ${-8.97}_{0.29}^{0.31}$ & ${0.15}_{0.11}^{0.10}$ & $86.56\pm3.37$ \\
O3ELG6 & $3.90\pm0.35$ & ${7.90}_{0.17}^{0.17}$ & --- & ${0.04}_{0.01}^{0.01}$ & ${0.01}_{0.00}^{0.00}$ & ${6.74}_{0.30}^{0.24}$ & ${-8.13}_{0.26}^{0.32}$ & ${0.09}_{0.06}^{0.04}$ & $377.78\pm52.77$ \\
O3ELG7 & $13.20\pm2.07$ & ${7.60}_{0.14}^{0.15}$ & --- & ${0.14}_{0.03}^{0.03}$ & ${0.02}_{0.00}^{0.00}$ & ${6.65}_{0.24}^{0.17}$ & ${-7.49}_{0.20}^{0.26}$ & ${0.02}_{0.02}^{0.01}$ & $1271.56\pm159.51^{*}$ \\
O3ELG8 & $4.78\pm0.69$ & ${7.89}_{0.19}^{0.19}$ & --- & ${0.02}_{0.01}^{0.01}$ & ${0.17}_{0.05}^{0.08}$ & ${7.79}_{0.23}^{0.22}$ & ${-9.40}_{0.26}^{0.29}$ & ${0.18}_{0.12}^{0.09}$ & $73.40\pm5.09$ \\
O3ELG9 & $6.22\pm0.75$ & ${7.60}_{0.14}^{0.15}$ & ${7.577}_{0.16}^{0.35}$ & ${1.25}_{0.28}^{0.29}$ & ${0.28}_{0.04}^{0.03}$ & ${7.69}_{0.26}^{0.15}$ & ${-7.60}_{0.18}^{0.28}$ & ${0.02}_{0.03}^{0.02}$ & $948.81\pm43.29$ \\
O3ELG10 & $5.28\pm3.61$ & ${7.75}_{0.29}^{0.33}$ & --- & ${0.00}_{0.00}^{0.00}$ & ${0.00}_{0.00}^{0.00}$ & ${6.31}_{0.28}^{0.28}$ & ${-8.85}_{0.33}^{0.37}$ & ${0.15}_{0.17}^{0.12}$ & $129.20\pm26.73$ \\
O3ELG11 & $5.10\pm1.43$ & ${7.92}_{0.17}^{0.16}$ & ${7.819}_{0.21}^{0.16}$ & ${0.34}_{0.15}^{0.15}$ & ${0.16}_{0.05}^{0.06}$ & ${7.73}_{0.37}^{0.27}$ & ${-8.19}_{0.33}^{0.42}$ & ${0.11}_{0.11}^{0.12}$ & $265.49\pm45.00$ \\
O3ELG12a & $2.80\pm0.71$ & ${7.88}_{0.23}^{0.22}$ & --- & ${0.25}_{0.11}^{0.14}$ & ${0.11}_{0.04}^{0.05}$ & ${7.61}_{0.35}^{0.29}$ & ${-8.22}_{0.35}^{0.43}$ & ${0.17}_{0.16}^{0.12}$ & $74.78\pm11.24$ \\
O3ELG12b & $8.71\pm0.38$ & ${7.53}_{0.15}^{0.16}$ & ${7.558}_{0.28}^{0.09}$ & ${0.05}_{0.01}^{0.01}$ & ${0.01}_{0.00}^{0.00}$ & ${6.68}_{0.31}^{0.24}$ & ${-7.99}_{0.26}^{0.33}$ & ${-0.01}_{0.07}^{0.05}$ & $718.65\pm73.90$ \\
O3ELG13 & $2.83\pm0.49$ & ${7.88}_{0.16}^{0.16}$ & --- & ${0.40}_{0.23}^{0.21}$ & ${0.44}_{0.11}^{0.19}$ & ${8.09}_{0.28}^{0.20}$ & ${-8.48}_{0.32}^{0.36}$ & ${0.16}_{0.15}^{0.16}$ & $237.72\pm16.17$ \\
O3ELG14 & $3.25\pm0.50$ & ${7.85}_{0.15}^{0.16}$ & --- & ${0.21}_{0.06}^{0.07}$ & ${0.08}_{0.01}^{0.01}$ & ${7.31}_{0.29}^{0.20}$ & ${-7.98}_{0.24}^{0.32}$ & ${0.04}_{0.04}^{0.03}$ & $351.67\pm51.42^{*}$ \\
O3ELG15 & $3.24\pm0.61$ & ${7.92}_{0.15}^{0.15}$ & --- & ${0.04}_{0.01}^{0.02}$ & ${0.07}_{0.02}^{0.02}$ & ${7.42}_{0.36}^{0.29}$ & ${-8.80}_{0.33}^{0.40}$ & ${0.16}_{0.10}^{0.08}$ & $202.69\pm35.35$ \\
O3ELG16 & $6.65\pm1.90$ & ${7.66}_{0.20}^{0.24}$ & --- & ${0.05}_{0.01}^{0.02}$ & ${0.01}_{0.00}^{0.00}$ & ${6.70}_{0.26}^{0.23}$ & ${-7.97}_{0.26}^{0.29}$ & ${0.09}_{0.06}^{0.05}$ & $114.67\pm13.07$
\enddata
\tablecomments{The log(O/H)+12 values are from the analysis discussed in section Section~\ref{sec:met}.  The log(O/H)+12 ($T_e$) values are the metallicity values inferred from the electron temperature direct method. The Log(M/$M_\odot$) and SFR(100Myr)($M_\odot$yr$^{-1})$) are from the SED fitting discussed in Section~\ref{sec:mass}. SFR(H$\alpha$)($M_\odot$yr$^{-1})$) are derived using the H$\alpha$ line flux as discussed in Section~\ref{sec:sfr}. The sSFR(yr$^{-1}$) values are derived from the H$\alpha$ SFR and the stellar mass derived from MCSED.} We lacked the high-quality photometry required to determine reliable SED fits for O3ELG10 so SED derived properties are estimates. \OIII$\lambda$5007 EWs measured from using the continuum in teh VIRUS spectra. In the few objects (indicated with an asterisk) where continuum was not significantly detected in the VIRUS spectra continuum levels are estimated from the MCSED fit. *Note O3ELG4b is not included in our sample as it did not meet our selection criteria but its properties are presented since it is a companion of O3ELG4a.
\end{deluxetable*}

\section{Individual Objects}
\label{sec:obj_dis}

With the limited number of objects in our study, and given that many are unique in terms of their properties, we provide a brief discussion of each below. VIRUS and LRS2 spectra, along with image cutouts are shown for each object in Figure~\ref{fig:spec_p1} and \ref{fig:spec_p2} in  Appendix~\ref{sec:data_panels}. \par

\subsection{O3ELG1}
\label{sec:1}
This object lies in the HETDEX fall field and represents a typical object for our sample. It has fairly low metallicity and mass, agreeing with the \citet{ber12} trend for typical low-mass galaxies. Like most objects in our sample it has a high SFR for its mass, similar to the blueberry galaxies. O3ELG1 is quite faint and has a compact morphology.

\subsection{O3ELG2}
\label{sec:2}
O3ELG2 is on the lower mass end for objects in the sample; however, it appears to be fairly typical of field galaxies in terms of its metallicity and mass. This situation is not the case for its SFR, as it falls in a similar regime as the blueberry galaxies. It appears to have an incredibly blue and compact morphology. This object, despite being fairly faint, has a rich emission-line spectrum that includes the \OIII $\lambda 4363$ line. This object also has the second highest EW in our sample overlapping with the lower end of the EW distribution of blueberry galaxies.

\subsection{O3ELG3}
\label{sec:3}
O3ELG3 is one of our faintest objects and has the second highest O3O2 ratio from its VIRUS spectrum. This object is our second lowest metallicity object and also one of the lowest mass objects. Like most of our objects, ours falls in the same regime as the typical low-mass galaxies in terms of its mass and metallicity. However, it has a high specific star formation rate similar to blueberry galaxies. It appears very blue and has a compact morphology.

\subsection{O3ELG4a+b}
\label{sec:4}

This object is one of the most interesting objects photometrically. The imaging shows two long, extended structures with a redder, compact object in between. The LRS2 follow-up confirmed that the two extended structures are galaxies at the same redshift, while the red compact object is likely a foreground star. The HETDEX detection found from our candidate selection was centered on O3ELG4a, however, since LRS2 follow up revealed O3ELG4b to be at the same redshift we determine properties of both out of interest. O3ELG4a is the highest mass galaxy included in our sample. It has low metallicity for its mass compared to typical field galaxies where its SFR is slightly higher than typical SDSS field galaxies however much more typical compared to our sample. O3ELG4b is not included in our final sample count since it did not meet our selection criteria and its properties reveal it has higher mass and is a fairly typical field galaxy in terms of its mass-metallicty and mass-SFR. It is interesting how much these two galaxies differ in properties, since it is likely they live in the same dark matter halo. 

\subsection{O3ELG5}
\label{sec:5}

This object was found with several independent detections in the HETDEX data; among these detections, it had one of the highest O3O2 ratios in sample.  However, the line ratios were not as high when the spectrum of the object was optimally extracted.  This is one of the highest metallicity and mass objects in our sample, but it still consistent with the mass-metallicity relation. However, it does have slightly higher specific star formation rate than a typical field galaxy at this redshift, falling at the lower end of the regime occupied by blueberry galaxies.

\subsection{O3ELG6}
\label{sec:6}

O3ELG6 is one of the lowest mass objects in our sample, with photometry revealing a faint compact morphology. This is another object found with multiple detections in HETDEX, some with very high O3O2 ratios, but others with more moderate ratios, and optimal extraction yields a more moderate ratio in combination. This object has a typical metallicity for our sample, and its low mass is typical in comparison to the \citet{and13} and \citet{ber12} star forming samples. O3ELG6 has one of the lowest SFR for the sample, but for its low mass it is much higher than typical field galaxies. 

\subsection{O3ELG7}
\label{sec:7}
O3ELG7 is by far our highest O3O2 ratio and \OIII $\lambda 5007$ EW object. It is also the lowest metallicity source without a  \OIII $\lambda 4363$ detection. O3ELG7 has one of the lowest masses and is also our highest redshift object at $z=0.0995$.  This object has a mass and metallicity consistent with the \citet{ber12} low-mass field galaxies. For its mass, this object is one of the most extreme in terms of its specific star formation rate. It clearly falls in a similar regime occupied by blueberry galaxies in terms of these two properties. It appears blue and compact in its morphology. Even though this object is consistent with typical field galaxies in terms of its mass and metallicity, at its very low mass this space is shared with a subset of the blueberry sample. In terms of its properties this object is the most similar to the blueberry population.

\subsection{O3ELG8}
\label{sec:8}

This object is near the high mass and metallicity end for our sample and has the second lowest SFR\null. In terms of its mass-metallicity and mass-SFR, O3ELG8 is one of the more typical galaxies in our sample; it is, however, still slightly high SFR for its mass. It is not as blue as other objects, but it still has a fairly compact morphology. 

\subsection{O3ELG9}
\label{sec:9}

This object is by far the brightest object and is the only object targeted for SDSS spectroscopy in our sample. Despite being quite extended in the images, it is less than 1$\sigma$ from the mean redshift of the sample, suggesting that it is likely more extended than most of our other objects. It is the third lowest metallicity object, but has one of the highest masses in the sample.
O3ELG9 has the highest SFR for our sample. The spectrum of O3ELG9 is rich with emission lines, including a significant \OIII $\lambda 4363$ detection. This object also has one of the highest \OIII~ EWs in our sample, overlapping the low end of the EW distribution for the blueberry galaxies.

\subsection{O3ELG10}
\label{sec:10}

Despite being the second lowest redshift object in the sample, O3ELG10 is the faintest object with g=21.9 and is almost undetected in the available imaging. This object has such a low S/N ratio VIRUS spectrum that it did not make the initial sample of strong candidates. However, it was among a group of promising fainter candidates and we were able to confirm its redshift with LRS2 spectra. This is one of two objects that does not have a GALEX detection, possibly due to its very low SFR (lowest in the sample). Unfortunately due to the faintness and limited photometric data available, we were not able to obtain reliable SED fits. However, since this object has the second lowest redshift and is also the faintest in photometry in our sample, we expect this is the lowest mass object. 
O3ELG10 has a metallicity typical for our sample and typical, if not high, for its mass in comparison with other field samples. This object has a SFR that falls between that of typical low mass galaxies and the more extreme blueberry galaxies.

\subsection{O3ELG11}
\label{sec:11}

O3ELG11 is one of the most extended objects in our sample and was found with multiple HETDEX detections. Its LRS2-B spectrum includes a significant \OIII $\lambda 4363$ detection, allowing us to infer metallicity from the direct $T_e$ method. 
This object has properties close to typical field galaxies in terms of its mass and metallicity. O3ELG11 does have high SFR for its mass, however, and falls in a similar SFR regime as the blueberry galaxies. 

\subsection{O3ELG12a+b}
\label{sec:12}

O3ELG12a and O3ELG12b were found with a single HETDEX detection. LRS2 observations confirmed the existence of two spatially separate sources at the same redshift. O3ELG12b has much higher O3O2 ratio than O3ELG12a, and hence a lower log(O/H)+12 by $\sim$0.2~dex and has lower stellar mass by $\sim$0.9~dex. O3ELG12b also appears physically much smaller in imaging; however, both have a compact morphology. Both have mass-metallicities consistent with the typical low-mass \citet{ber12} sample but high SFRs for their mass. This object is an interesting case, because the two components are sufficiently close to imply an interaction, but exhibit low but different metallicities, with O3ELG12b much less massive. Interestingly O3ELG12b has a much higher EW consistent with the less extreme end of the blueberry sample where O3ELGa has an EW typical of our sample and generally lower than the blueberries. The disparity in mass and metallicities between the galaxies is shared with the O3ELG4a and b galaxy pair.

\subsection{O3ELG13}
\label{sec:13}

O3ELG13 was a strong candidate, with a moderately high O3O2 ratio in its HETDEX spectrum. This object is one of the highest mass objects in the sample and falls in the regime typical of low-mass dwarfs in terms of its mass and metallicity. It does, however, have a fairly high SFR for its mass.  Like most of the objects in the sample it has a blue, compact morphology.

\subsection{O3ELG14}
\label{sec:14}

O3ELG14 has a moderately high O3O2 ratio in its HETDEX spectrum, and its redshift, $z=0.0904$, is the second highest in our sample. It is fairly typical in terms of its mass and metallicity. It does, however, have one of the highest SFR and has a high specific SFR.  

\subsection{O3ELG15}
\label{sec:15}

O3ELG15 is an extended source in imaging data. It is also the lowest redshift object in the sample at $z=0.0267$ and has a spectrum rich in emission lines. Since its  \OIII $\lambda 4363$ line is well-detected, the direct $T_e$ method was employed to infer its metallicity. Its log(O/H)+12=7.45 makes it the lowest metallicity object with an \OIII $\lambda 4363$ detection. This object has typical metallicity for its mass consistent with the \citet{ber12} sample. It is still highly star forming for its mass, however, it is one of the more typical objects in terms of its SFR for our sample falling in the very low end of specific star formation rates of blueberries.

\subsection{O3ELG16}
\label{sec:16}

O3ELG16 has one of the highest O3O2 ratios and is one of the lowest mass objects in the sample. This object is also one of our faintest objects and is challenging to observe due to its proximity to a bright star on-sky. This is one of two objects that does not have a GALEX detection either due to it being a faint UV source or its proximity to the bright star. 
It has metallicity fairly typical for its low mass; however, it is very highly star forming for its mass compared to typical field galaxies. 

\section{Discussion}
\label{sec:disussion}

In terms of the mass and metallicity our galaxies, for the most part, are consistent with the typical low-mass dwarf galaxies found by \citet{ber12}. However, our sample has high specific SFRs similar to the blueberry sample. Our objects have high \OIII ~EWs, however, for the majority of our sample, they are not nearly as extreme as the \OIII ~EWs of the  blueberry galaxies. With spectroscopic selection we find a sample of galaxies that is a hybrid between the properties of blueberry galaxies and typical low-mass dwarfs.

This result is not unexpected as photometric selection is less sensitive to finding very low mass, and hence faint continuum emission from older stellar populations, star forming galaxies unless they have extreme EW emission that can be picked up in broad band filters. The first internal HETDEX data release has allowed us to find a significant sample over a relatively small volume of very low-mass, star forming galaxies at low redshift that fall near or below the photometric detection limits of SDSS and many other wide field photometric surveys. Our sources have a faint continuum and only our brightest object (O3ELG9) was selected for follow up spectroscopy in SDSS. The faintest source in our sample (O3ELG10) does not even have a photometry match in SDSS.

The low density of extreme EW \OIII $\lambda 5007$ galaxies found in broad band surveys implies that these objects are incredibly rare.  Only 251 green peas ($0.112 < z < 0.360$; \citealt{car09}) and 40 blueberry galaxies ($z<0.05$; \citealt{yan17}) have been identified in the SDSS\null. Based on this number density for blueberry galaxies ($3.0 \times 10^{-6}$~Mpc$^{-3}$; \citealt{yan17}) we should not have expected to find any of these objects in all of HETDEX\null. However, we easily identified a significant sample of galaxies with similar mass, SFR, and similar to slightly higher metallicity as blueberries in the first $\sim 9$\% of HETDEX observations.  ``Easily'' here refers to the conservative search criteria used to build our sample. Our sample of objects is not the same as the blueberry sample, however, our spectroscopically-selected sample is complementary as it finds objects with more intermediate \OIII ~EW values that are missed by these broad band filter studies. We expect many fainter galaxies and possibly even higher O3O2 ratio sources exist in the full HETDEX data set. \par

Since we do not attempt to build a complete sample in this pilot study, we cannot predict a final number density for these sources in HETDEX; we can, however, provide a strong lower limit. The density of objects with high O3O2 ratios is $>$2.08 objects per deg$^2$ and $>$2.8e-4 per Mpc$^3$. This pilot HETDEX study finds $\sim$250 more galaxies per volume sampled than blueberry galaxies found through SDSS photometry \citep{yan17}, showing our objects are much less rare.\par

Table~\ref{comp_tbl} summarizes the comparison studies of the populations of low-redshift populations discussed in Section~\ref{sec:intro} and shown in Figure~\ref{fig:massmet} and \ref{fig:sfr}. The surface density and volume density estimates are calculated from the number of objects found over the search area and redshift range. The luminous compact objects were not included for comparison as they are similar objects to green peas but extend to much higher redshift. The KISS survey finds slightly more objects per deg$^2$ and per Mpc$^3$ than the broad band \citet{car09} study. The \citet{kak07} and the \citet{hu09} ultra-strong emission-line galaxies (USELG) sample revealed seven galaxies they classify as extremely metal poor galaxies (XMPG) with log(O/H)+12 $<$ 8.65. There are 21 additional \OIII $\lambda$5007 galaxies in their sample that have 7.65 $<$ log(O/H)+12 $<$ 8.0. This sample does illustrate that narrow-band imaging is effective at finding low metallicity systems.\par 

\begin{deluxetable*}{cccccc}[ht!]
\tablecaption{Properties of surveys for faint nearby galaxies \label{comp_tbl}}
\tablehead{\colhead{Sample} & \colhead{Selection} & \colhead{Sample Size}  & \colhead{redshift range} & \colhead{Surface Density} & \colhead{Volume Density} \\
\colhead{} & \colhead{} & \colhead{}  & \colhead{} & \colhead{(deg${}^{-2}$)} & \colhead{(Mpc${}^{-3}$)}}
\startdata
\textbf{HETDEX O3ELG Sample} & Spectroscopy & 17 & 0.0-0.1 & $>$2.08 & $>$2.8e-4 \\
\textbf{HETDEX Pilot Survey [1]} & Spectroscopy & 2* & 0.0-0.15 & 44.17 & 1.8e-3 \\
\textbf{Blue Compact Dwarfs [2]} & Broad Band & 2023 & 0.0-0.05 & 0.139 & 1.44e-4 \\
\textbf{Green Peas [3]} & Broad Band & 251 & 0.112-0.360 & 0.0298 & 1.084e-7 \\
\textbf{Blueberries [4]} & Broad Band & 43 & 0.0-0.05 & 0.00295 & 1.137e-6 \\
Luminous Compact Objects [5] & Broad Band & 803 & 0.2-0.63 & 0.0953 & 7.737e-8 \\
KISS Green Peas [6] & Objective Prism & 15 & 0.29-0.41 & 0.100 & 4.051e-7 \\
USELG [7]  & Narrow Band & 28 & 0.616-0.640, 0.813-0.837 & 56.0 & 1.13e-4
\enddata
\tablecomments{[1] \citet{ind19}, [2] \citet{lia16}, [3] \citet{car09}, [4] \citet{yan17}, [5] \citet{izo11}, [6] \citet{bru20}, [7] Ultra-strong emission line galaxies with log(O/H)+12$<$8.0; \citet{kak07, hu09}. The bold face samples are represented in Figure~\ref{fig:massmet} for comparison. *Note the HETDEX Pilot Survey is only counting the two low-mass objects found in the sample.}
\end{deluxetable*}

The drawbacks of narrow-band imaging are the narrow redshift range accessible to any given filter, and the need for subsequent spectroscopic observations to confirm the nature of the objects.  This latter requirement can be quite expensive in terms of telescope time. The \citet{ind19} HPS investigation, as well as this pilot study, shows that wide-field IFU spectroscopy is much more sensitive to finding low metallicity, continuum-faint galaxies than broad band imaging.  While the HPS study found a much higher surface and volume density of \OIII $\lambda$5007 galaxies in their attempt to build a complete sample of \OII and \OIII\ emitters over a small volume, the current work searched only for galaxies with high O3O2 ratios over a much larger volume of the universe and hence finds a much lower density of these low-mass, low-metallicity objects. \par

By assembling a sample of $\sim$20 extremely low metallicity galaxies from many studies in the literature \citet{mcq20} suggests that galaxies that deviate in their luminosity-mass relation are likely experiencing an episode of intense star formation. Based on their analysis it seems likely that the galaxies in our sample are following normal secular evolution pathways but are experiencing a recent episode of star formation. For the blueberry galaxies that also deviate from the mass-metallicity relation, the recent star formation is likely triggered by an outside interaction that has resulted in infalling, low-metallicity gas. Since the \citet{ber12} galaxies are consistent with both the mass-luminosity and mass-metallicity relations these objects are following normal secular evolution pathways and are low metallicity simply due having gravitational potential due to their low mass allowing for outflows from past star formation events to more easily escape the galaxy \citep{mcq20}. 

Using the lower limit number density of our sample of galaxies with high specific star formation rates and the \citet{ber12} sample of relatively inactive galaxies we can estimate a lower limit on the duty cycle of galaxies at low redshift. From the Local Volume Legacy (LVL) Survey 42 galaxies are identified as low luminosity \citep{ber12}. Even though \citet{ber12} was only able to get direct ($T_e$) metallicity estimates for 31 of these objects we assume the complete low luminosity sample of the nearest 11Mpc have similar properties. We estimate a number of typical dwarf galaxies in the iHDR1 survey volume to be $\sim$524. With the caveats of the extrapolation of the LVL number density of low luminosity galaxies (z$\lesssim$0.002455) to our higher redshift range (z$\lesssim$0.1) we estimate a lower limit on the low-redshift galaxy duty cycle to be $\sim$3.14$\%$ of galaxies are experiencing an episode of intense star formation.\par

Our study has also shown that LRS2 follow-up of VIRUS-detected \OIII $\lambda$5007 galaxies will not be necessary in the larger HETDEX survey going forward, so long as multiple line detections can be identified. The addition of the red LRS2 lines did not significantly change the values derived in the metallicity fits. These fits can be made on blue lines within the VIRUS bandpass, although with larger uncertainties. VIRUS has a very blue wavelength limit of $\lambda$=3500\AA\ compared to typical IFU instruments and is extremely sensitive in finding low-redshift, low metallicity galaxies. \par 

This pilot study utilizing early HETDEX data paves the way for building a large sample of low-mass, low-metallicity galaxies at low-redshift with the complete HETDEX survey. Based on the predicted lower-limit volume density derived above, we expect to find at least $\sim$250 objects with properties similar to those in the current sample. This larger sample would provide better statistics to build an understanding of galaxy properties that bridge the gap in the mass-metallicity-SFR phase space that spans blueberries and typical dwarf galaxies like those in the \citep{ber12} sample. 
Likely there are many more \OIII $\lambda 5007$ emitters with S/N $<$ 10 and objects with more extreme O3O2 ratios such that the \OII $\lambda 3727$ lines is undetected, left out of our sample. By relaxing our conservative search criteria in building a larger sample, many more objects could be found that are either fainter or with more extreme O3O2 ratios. O3ELG10 is an example of such an object. \par

Finally, this study confirms that blind spectroscopy without any photometric pre-selection provides an excellent tool for finding metal-poor populations of galaxies that are not easily identified in continuum-based surveys. Our sources on average tend to be typical dwarf galaxies in their mass-metallicity properties, but are however extreme in the mass-SFR regime. The complete HETDEX survey will provide spectra for a large sample of these actively star forming dwarf galaxies at low-redshift. This larger sample will allow for statistical studies to be made with a unique sample of galaxies.  \par

\section{Conclusions} 
\label{sec:conclusions}

We assembled a sample of spectroscopically-selected, low-metallicity objects at $z<0.1$ from an early internal HETDEX data release covering 8.16~deg$^2$. Using conservative selection criteria, i.e., requiring the S/N of \OIII $\lambda 5007$ to be greater than 10 and the VIRUS detection spectrum to have at least three distinguishable emission lines, 17 sources of interest were identified with high O3O2 ratios. All of these objects were re-extracted from the VIRUS data and also followed up with HET LRS2-R spectroscopy to measure H$\alpha$. LRS2-B spectroscopy was also obtained for all but one of the sample. \par

Properties linked to the evolutionary history of these objects, such as stellar mass, metallicity, and instantaneous SFR, and \OIII $\lambda 5007$ EW were measured. A comparison of our sample with samples of photometrically selected ``typical'' and ``extreme'' galaxies (such as blueberries and green peas), reveals that our objects have, on average, lower metallicity and higher SFR compared to typical SDSS objects. However, our study finds objects with much lower masses and we conclude are still consistent with the mass-metallicity relation when compared with typical low-mass dwarf galaxies from \citet{ber12}. Interestingly, these objects have much higher sSFRs than the \citet{ber12} sample. In comparing their \OIII EW values to that of blueberry galaxies, our sample has high EW values, however, for the most part not as extreme as the blueberry sample. This study illustrates that selecting low-redshift galaxies purely based on their oxygen emission is successful at finding low mass dwarfs that are much more actively star forming on average for their mass. With the complete HETDEX survey we will be able to build a large sample of these objects, allowing them to be studied statistically to provide a more complete picture of low redshift galaxy evolution.

\acknowledgments
HETDEX is led by the University of Texas at Austin McDonald Observatory and Department of Astronomy with participation from the Ludwig-Maximilians-Universit\"at M\"unchen, Max-Planck-Institut f\"ur Extraterrestriche Physik (MPE), Leibniz-Institut f\"ur Astrophysik Potsdam (AIP), Texas A\&M University, the Pennsylvania State University, Institut f\"ur Astrophysik G\"ottingen, the University of Oxford, Max-Planck-Institut f\"ur Astrophysik (MPA), the University of Tokyo, and Missouri University of Science and Technology. In addition to Institutional support, HETDEX is funded by the National Science Foundation (grant AST-0926815), the State of Texas, the US Air Force (AFRL FA9451-04-2-0355), and generous support from private individuals and foundations.\par

The observations were obtained with the Hobby-Eberly Telescope (HET), which is a joint project of the University of Texas at Austin, the Pennsylvania State University, Ludwig-Maximilians-Universität M\"unchen, and Georg-August- Universit\"at G\"ottingen. The HET is named in honor of its principal benefactors, William P. Hobby and Robert E. Eberly.\par

The authors acknowledge the Texas Advanced Computing Center (TACC) at The University of Texas at Austin for providing high performance computing, visualization, and storage resources that have contributed to the research results reported within this paper. URL: http://www.tacc.utexas.edu \par

We thank the staff of the Hobby Eberly Telescope (HET) for support of our LRS2 observations as well as the LRS2 commissioning team. The LRS2 spectrograph was funded by McDonald Observatory and Department of Astronomy, University of Texas at Austin, and the Pennsylvania State University. We thank the HET board for allocating LRS2 GTO time to the LRS2 commissioning team as some of this time was used for our LRS2 follow-up. We thank the Institute for Gravitation and the Cosmos which is supported by the Eberly College of Science and the Office of the Senior Vice President for Research at the Pennsylvania State University.

\vspace{5mm}
Facilities: \facilities{HET(LRS2), HET(VIRUS), TACC}

\software{\texttt{astropy} \citep{ast13, ast18},  
          \texttt{emcee} \citep{for13}, 
          \texttt{LA Cosmics} \citep{van01}
          \texttt{\texttt{MCSED}} (\citep{bow20})}

%\newpage
\appendix
\label{appendix}

\section{Data Panels}
\label{sec:data_panels}
\begin{figure*}[ht!]
\epsscale{0.9}
\plotone{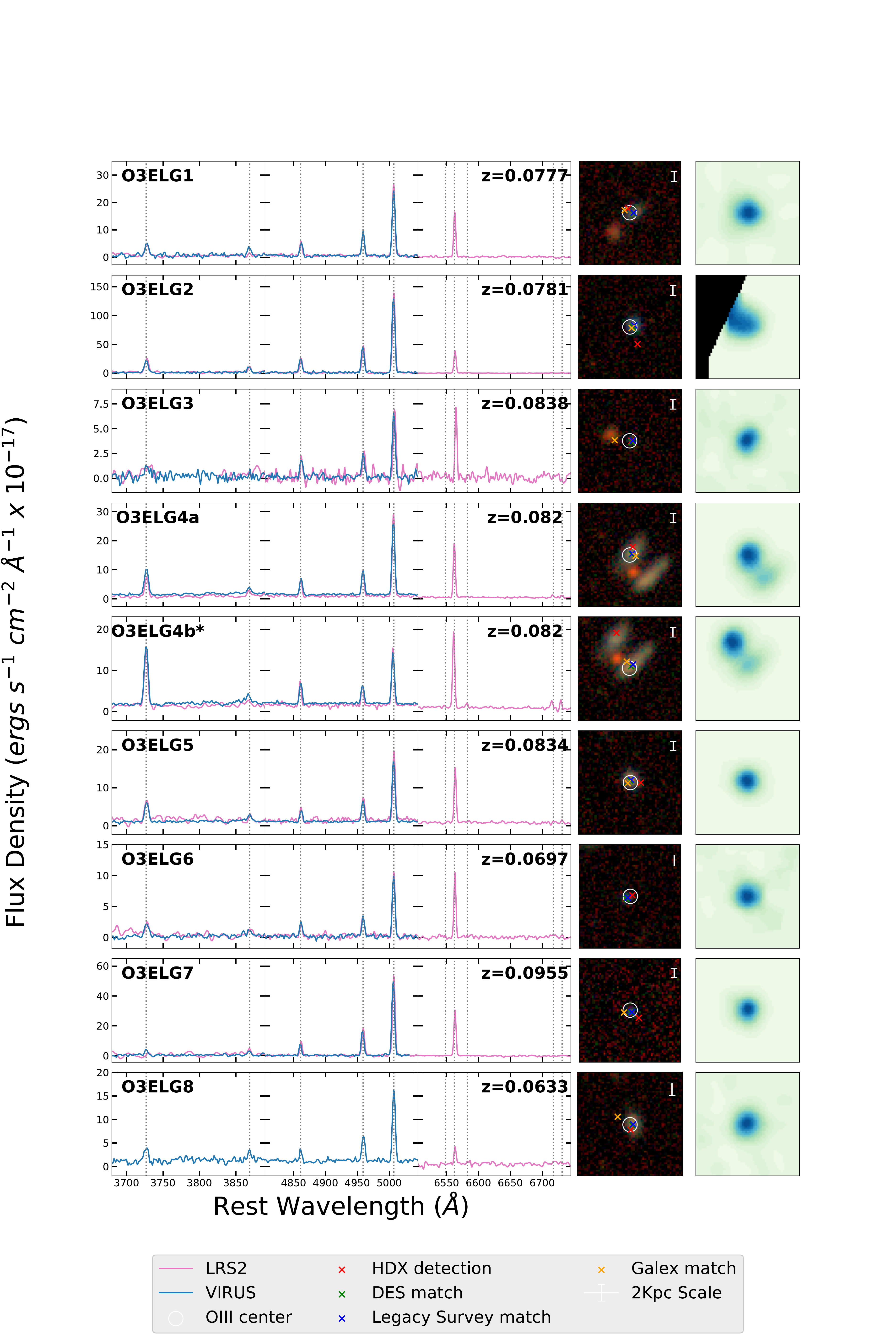}
\caption{Panels showing the VIRUS and LRS2 spectra on the left, DESI Legacy Imaging Surveys color cutouts with photometry matches in the middle, and VIRUS \OIII $\lambda 5007$ maps on the right for objects O3ELG1-8.}
\label{fig:spec_p1}
\end{figure*}

\begin{figure*}[ht!]
\epsscale{0.9}
\plotone{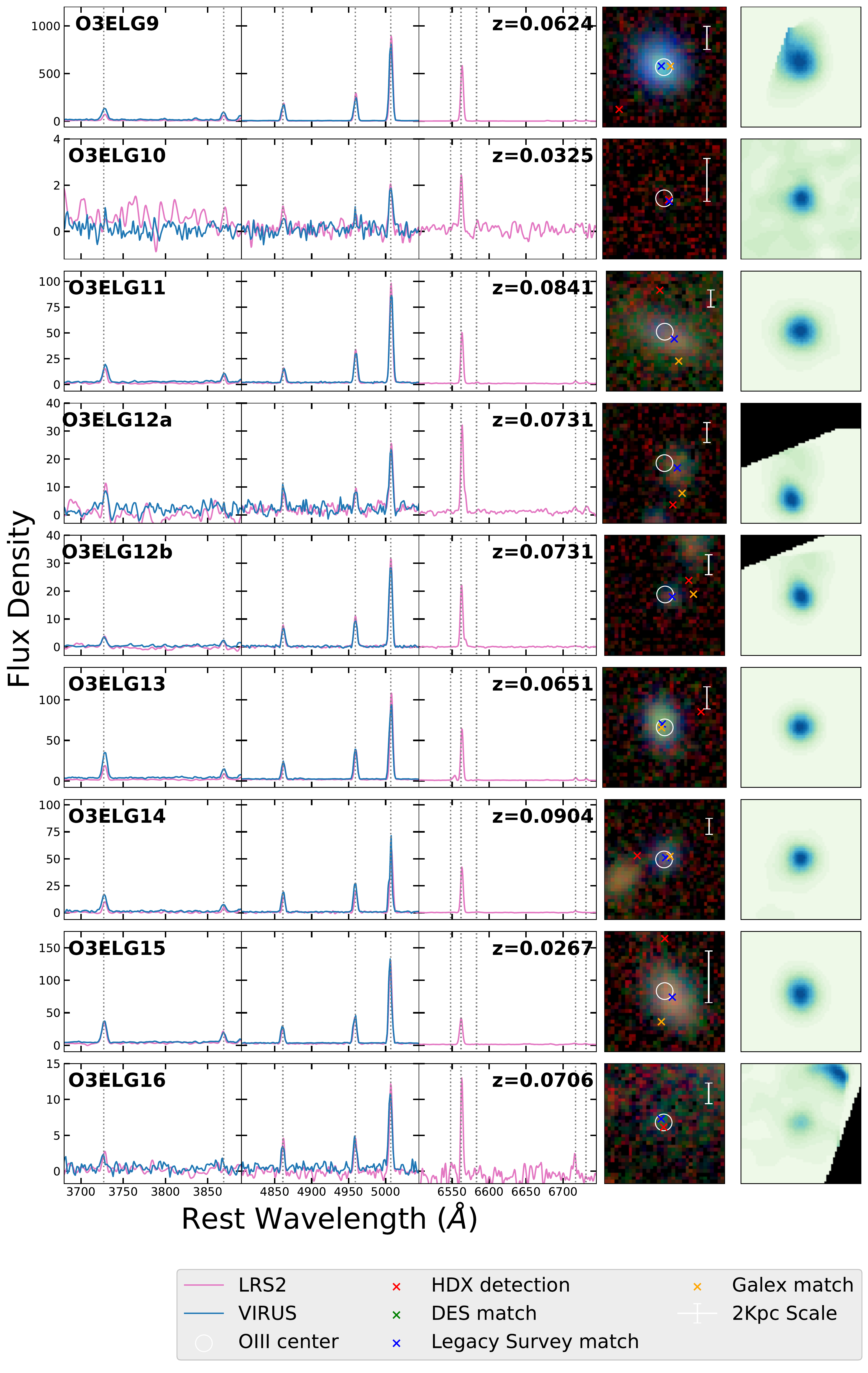}
\caption{Panels showing the VIRUS and LRS2 spectra on the left, DESI Legacy Imaging Surveys color cutouts with photometry matches in the middle, and VIRUS \OIII $\lambda 5007$ maps on the right for objects O3ELG9-16.}
\label{fig:spec_p2}
\end{figure*}

The panels in Figure~\ref{fig:spec_p1} and \ref{fig:spec_p2} show the HETDEX (VIRUS) and LRS2 data for each of the 17 objects in the sample and also O3ELG4b. The leftmost panels show spectra from VIRUS (blue), the LRS2 (pink) in three wavelength regimes and highlights the primary emission lines used in this study. Refer to Figure~\ref{fig:lrs2_spec} for a few examples of complete LRS2 spectra.  The LRS2 data have been scaled to the VIRUS spectra using the overlapping \OIII $\lambda 5007$ line, except in the case of O3ELG8, where we lack an LRS2-B spectrum to provide a scale factor.\par

The panels on the far right present $14\arcsec \times 14\arcsec$ maps of \OIII $\lambda 5007$ constructed from the VIRUS data. The hard cutoff boundaries in some of these images, for example in O3ELG2, 9, and 12a+b, are due to the object being located on the edge of a field. The middle panel displays $14\arcsec \times 14\arcsec$ color ($g$,$r$,$z$) cutouts from the Legacy Survey. The white circle represents the centroid of the object in the \OIII $\lambda 5007$ VIRUS map (shown in the far right panel). The red X represents the position of the highest O3O2 ratio detection by HETDEX\null. The green, blue, and orange Xs represent the photometry catalog matches from the Dark Energy Survey (DES), the Legacy Survey, and GALEX, respectively. These cutouts have a scale bar in the upper right corner representing a 2~kpc scale at the redshift of the object.   \par

*Note O3ELG4b is not included in our sample as it did not meet our selection criteria but it is presented here since it is a companion of O3ELG4a. Looking at its spectrum it is clear that the O3O2 ratio of this object is much smaller than all of the galaxies in our sample. It is also very faint in its \OIII $\lambda 5007$ VIRUS map compared to O3ELG4a. \par

\pagebreak

\section{Best Fit SEDs}
\label{sec:sed_panels}

\begin{figure*}[ht!]
\epsscale{1.0}
\plotone{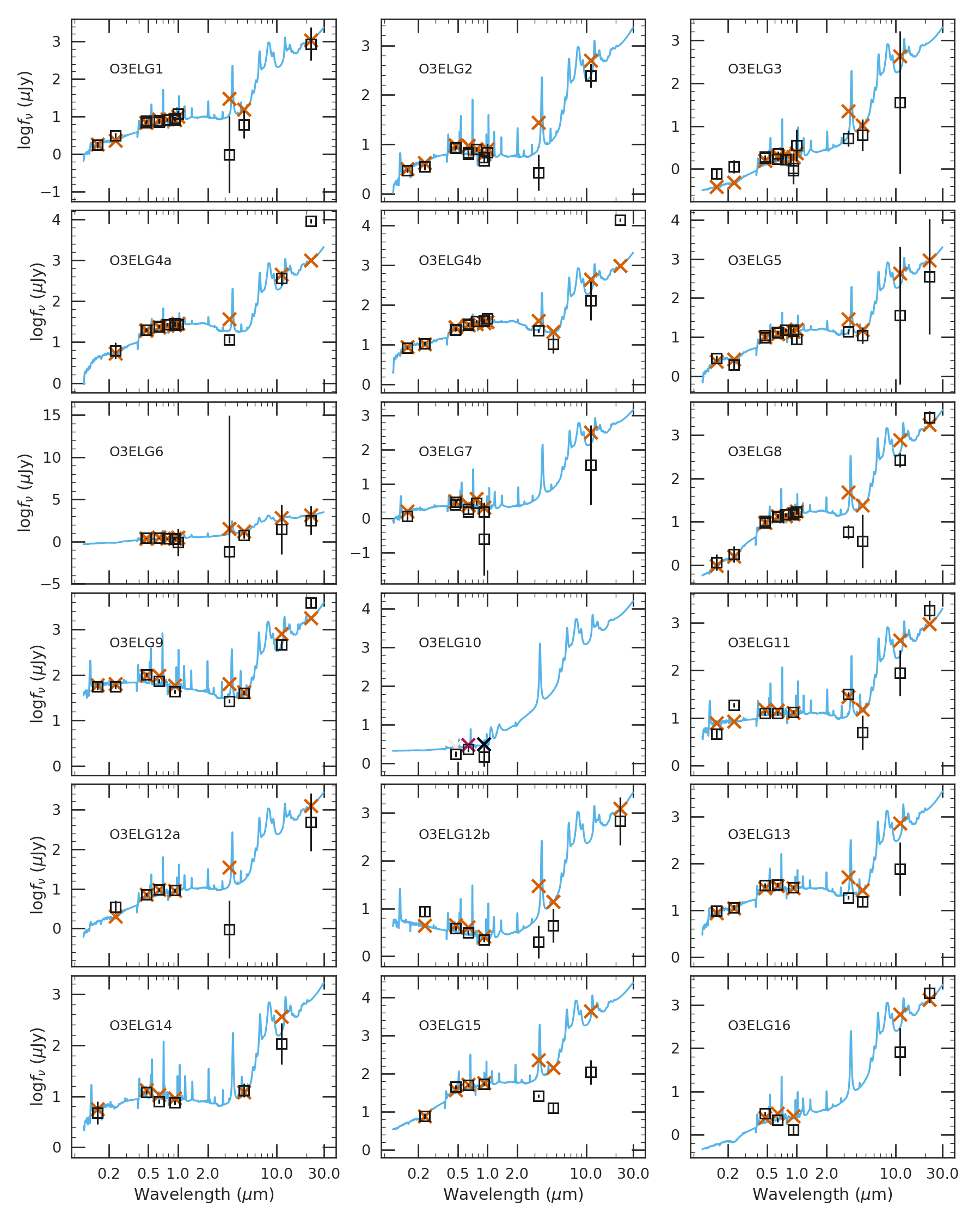}
\caption{The panels show the best fit SEDs (blue line),
predicted photometric fluxes (orange Xs),
and measured photometry (black squares) for the objects in our sample. The mismatch sometimes seen between the SED fits and the WISE bands is likely due to the fact that our models do not fully explore the PAH parameter space.}
\label{fig:sed_fits}
\end{figure*}

In panel \ref{fig:sed_fits} we present the best fit SEDs for each objects in our sample. The black squares represent the photometry and associated errors (presented in Section \ref{sec:photo}). The blue line
is the best fit SED model to the observed photometry for each source, and the orange Xs
represent the fits at each photometric data point. 

\pagebreak

\bibliographystyle{aasjournal}

\begin{thebibliography}{}

\bibitem[Abazajian et al.(2009)]{aba09} Abazajian, K.~N., Adelman-McCarthy, J.~K., Ag{\"u}eros, M.~A., et al.\ 2009, \apjs, 182, 543

\bibitem[Abbott et al.(2018)]{abb18} DES Collaboration, et al.\ 2018, \apjs, 239, 18

\bibitem[Adams et al.(2011)]{ada11} Adams, J.~J., Blanc, G.~A., Hill, G.~J., et al.\ 2011, \apjs, 192, 5

\bibitem[Alam et al.(2015)]{ala15} Alam, S., Albareti, F.~D., Allende Prieto, C., et al.\ 2015, \apjs, 219, 12

\bibitem[Allington-Smith et al.(2002)]{all02} Allington-Smith, J., Murray, G., Content, R., et al.\ 2002, \pasp, 114, 892

\bibitem[Amor{\'\i}n et al.(2010)]{amo10} Amor{\'\i}n, R.~O., P{\'e}rez-Montero, E., \& V{\'\i}lchez, J.~M.\ 2010, \apjl, 715, L128

\bibitem[Andrews \& Martini(2013)]{and13} Andrews, B.~H., \& Martini, P.\ 2013, \apj, 765, 140 

\bibitem[Astropy Collaboration et al.(2018)]{ast18} Astropy Collaboration, Price-Whelan, A.~M., Sip{\H o}cz, B.~M., et al.\ 2018, \aj, 156, 123 

\bibitem[Astropy Collaboration et al.(2013)]{ast13} Astropy Collaboration, Robitaille, T.~P., Tollerud, E.~J., et al.\ 2013, \aap, 558, A33 

\bibitem[Battisti et al.(2016)]{bat16} Battisti, A.~J., Calzetti, D., \& Chary, R.-R.\ 2016, \apj, 818, 13

\bibitem[Berg et al.(2012)]{ber12} Berg, D.~A., Skillman, E.~D., Marble, A.~R., et al.\ 2012, \apj, 754, 98 

\bibitem[Bertelli \etal(1994)]{bertelli+94} Bertelli, G., Bressan, A., Chiosi, C., \etal\ 1994, \aaps, 106, 275

%\bibitem[Bian et al.(2016)]{bia16} Bian, F., Kewley, L.~J., Dopita, M.~A., et al.\ 2016, \apj, 822, 62

%\bibitem[Bian et al.(2017)]{bia17} Bian, F., Kewley, L.~J., Dopita, M.~A., et al.\ 2017, \apj, 834, 51

\bibitem[Bian et al.(2018)]{bia18} Bian, F., Kewley, L.~J., \& Dopita, M.~A.\ 2018, \apj, 859, 175 

\bibitem[Blanc et al.(2013)]{bla13} Blanc, G.~A., Weinzirl, T., Song, M., et al.\ 2013, \aj, 145, 138

\bibitem[Bowman et al.(2020)]{bow20} Bowman, W.~P., Zeimann, G.~R., Nagaraj, G., et al.\ 2020, arXiv e-prints, arXiv:2006.13245

\bibitem[Brammer et al.(2012)]{bra12} Brammer, G.~B., van Dokkum, P.~G., Franx, M., et al.\ 2012, \apjs, 200, 13 

\bibitem[Brinchmann et al.(2004)]{bri04} Brinchmann, J., Charlot, S., White, S.~D.~M., et al.\ 2004, \mnras, 351, 1151

\bibitem[Brunker et al.(2020)]{bru20} Brunker, S.~W., Salzer, J.~J., Janowiecki, S., et al.\ 2020, arXiv e-prints, arXiv:2006.14663

\bibitem[Bryant et al.(2015)]{bry15} Bryant, J.~J., Owers, M.~S., Robotham, A.~S.~G., et al.\ 2015, \mnras, 447, 2857 

\bibitem[Bryant et al.(2016)]{bry16} Bryant, J.~J., Bland-Hawthorn, J., Lawrence, J., et al.\ 2016, \procspie, 9908, 99081F 

\bibitem[Bundy et al.(2015)]{bun15} Bundy, K., Bershady, M.~A., Law, D.~R., et al.\ 2015, \apj, 798, 7 

\bibitem[Byler \etal(2017)]{byler+17} Byler, N., Dalcanton, J.~J., Conroy, C., \etal\  2017, \apj, 840, 44

\bibitem[Calzetti et al.(2000)]{cal00} Calzetti, D., Armus, L., Bohlin, R.~C., et al.\ 2000, \apj, 533, 682 

\bibitem[Calzetti(2001)]{cal01} Calzetti, D.\ 2001, \pasp, 113, 1449

\bibitem[Cardamone et al.(2009)]{car09} Cardamone, C., Schawinski, K., Sarzi, M., et al.\ 2009, \mnras, 399, 1191 

%\bibitem[Cair{\'o}s et al.(2010)]{cai10} Cair{\'o}s, L.~M., Caon, N., Zurita, C., et al.\ 2010, \aap, 520, A90

\bibitem[Chabrier(2003)]{cha03} Chabrier, G.\ 2003, \pasp, 115, 763

\bibitem[Chonis et al.(2016)]{cho16} Chonis, T.~S., Hill, G.~J., Lee, H., et al.\ 2016, \procspie, 9908, 99084C 

\bibitem[Conroy et al.(2009)]{fsps-1} Conroy, C., Gunn, J.~E., \& White, M.\ 2009, \apj, 699, 486

\bibitem[Conroy \& Gunn(2010)]{fsps-2} Conroy, C., \& Gunn, J.E. 2010, \apj, 712, 833

\bibitem[Conroy(2013)]{con13} Conroy, C.\ 2013, \araa, 51, 393. doi:10.1146/annurev-astro-082812-141017

%\bibitem[Conselice(2014)]{con14} Conselice, C.~J.\ 2014, \araa, 52, 291

\bibitem[Curti et al.(2017)]{cur17} Curti, M., Cresci, G., Mannucci, F., et al.\ 2017, \mnras, 465, 1384 

\bibitem[Dale et al.(2009)]{dal09} Dale, D.~A., Cohen, S.~A., Johnson, L.~C., et al.\ 2009, \apj, 703, 517

\bibitem[Denicol{\'o} et al.(2002)]{den02} Denicol{\'o}, G., Terlevich, R., \& Terlevich, E.\ 2002, \mnras, 330, 69

\bibitem[Dey et al.(2019)]{dey19} Dey, A., Schlegel, D.~J., Lang, D., et al.\ 2019, \aj, 157, 168

\bibitem[Drory et al.(2001)]{dro01} Drory, N., Feulner, G., Bender, R., et al.\ 2001, \mnras, 325, 550 

\bibitem[Duarte Puertas et al.(2017)]{dua17} Duarte Puertas, S., Vilchez, J.~M., Iglesias-P{\'a}ramo, J., et al.\ 2017, \aap, 599, A71 

\bibitem[Ferland \etal(1998)]{CLOUDY} Ferland, G.J., Korista, K.T., Verner, D.A., \etal\  1998, \pasp, 110, 761 

\bibitem[Ferland \etal(2013)]{CLOUDY13} Ferland, G.J., Porter, R.L., van Hoof, P.A.M., \etal\  2013, Rev.~Mex.A.A., 49, 137

\bibitem[Foreman-Mackey et al.(2013)]{for13} Foreman-Mackey, D., Hogg, D.~W., Lang, D., \& Goodman, J.\ 2013, \pasp, 125, 306 

\bibitem[Gallego et al.(1996)]{gal96} Gallego, J., Zamorano, J., Rego, M., Alonso, O., \& Vitores, A.~G.\ 1996, \aaps, 120, 323 

\bibitem[Girardi \etal(2000)]{girardi+00} Girardi, L., Bressan, A., Bertelli, G., \& Chiosi, C. 2000, \aaps, 141, 371 

\bibitem[Grasshorn Gebhardt et al.(2016)]{gra16} Grasshorn Gebhardt, H.~S., Zeimann, G.~R., Ciardullo, R., et al.\ 2016, \apj, 817, 10 

\bibitem[Hawley(2012)]{haw12} Hawley, S.~A.\ 2012, \pasp, 124, 21 

%\bibitem[Henry et al.(2015)]{hen15} Henry, A., Scarlata, C., Martin, C.~L., et al.\ 2015, \apj, 809, 19

\bibitem[Hill et al.(2008a)]{hil08b} Hill, G.~J., Gebhardt, K., Komatsu, E., et al.\ 2008, Panoramic Views of Galaxy Formation and Evolution, 399, 115

\bibitem[Hill et al.(2008b)]{hil08a} Hill, G.~J., MacQueen, P.~J., Smith, M.~P., et al.\ 2008, \procspie, 7014, 701470

\bibitem[Hill \& HETDEX Consortium(2016)]{hil16} Hill, G.~J. \& HETDEX Consortium\ 2016, Multi-Object Spectroscopy in the Next Decade: Big Questions, Large Surveys, and Wide Fields, 507, 393

\bibitem[Hill et al.(2018)]{hil18a} Hill, G.~J., Kelz, A., Lee, H., et al.\ 2018, \procspie, 10702, 107021K

\bibitem[Hu et al.(2009)]{hu09} Hu, E.~M., Cowie, L.~L., Kakazu, Y., et al.\ 2009, \apj, 698, 2014

\bibitem[Indahl et al.(2019)]{ind19} Indahl, B., Zeimann, G., Hill, G.~J., et al.\ 2019, \apj, 883, 114

%\bibitem[Isobe et al.(2020)]{iso20} Isobe, Y., Ouchi, M., Kojima, T., et al.\ 2020, arXiv:2004.11444

%\bibitem[Izotov et al.(2018)]{izo18} Izotov, Y.~I., Worseck, G., Schaerer, D., et al.\ 2018, \mnras, 478, 4851

\bibitem[Izotov et al.(2006)]{izo06} Izotov, Y.~I., Stasi{\'n}ska, G., Meynet, G., Guseva, N.~G., \& Thuan, T.~X.\ 2006, \aap, 448, 955 

\bibitem[Izotov et al.(2011)]{izo11} Izotov, Y.~I., Guseva, N.~G., \& Thuan, T.~X.\ 2011, \apj, 728, 161 

\bibitem[Kakazu et al.(2007)]{kak07} Kakazu, Y., Cowie, L.~L., \& Hu, E.~M.\ 2007, \apj, 668, 853

%\bibitem[Kauffmann et al.(2003)]{kau03} Kauffmann, G., Heckman, T.~M., White, S.~D.~M., et al.\ 2003, \mnras, 341, 33 

\bibitem[Kennicutt(1998)]{ken98} Kennicutt, R.~C., Jr.\ 1998, \araa, 36, 189 

\bibitem[Kennicutt et al.(2008)]{ken08} Kennicutt, R.~C., Lee, J.~C., Funes, J.~G., et al.\ 2008, \apjs, 178, 247. doi:10.1086/590058

\bibitem[Kennicutt et al.(2009)]{ken09} Kennicutt, R.~C., Hao, C.-N., Calzetti, D., et al.\ 2009, \apj, 703, 1672

\bibitem[Kennicutt \& Evans(2012)]{ken12} Kennicutt, R.~C., \& Evans, N.~J.\ 2012, \araa, 50, 531

\bibitem[Kewley \& Dopita(2002)]{kew02} Kewley, L.~J., \& Dopita, M.~A.\ 2002, \apjs, 142, 35 

%\bibitem[Kewley et al.(2004)]{kew04} Kewley, L.~J., Geller, M.~J., \& Jansen, R.~A.\ 2004, \aj, 127, 2002 

\bibitem[Kewley \& Ellison(2008)]{kew08} Kewley, L.~J., \& Ellison, S.~L.\ 2008, \apj, 681, 1183 

%\bibitem[Kim et al.(2020)]{kim20} Kim, K., Malhotra, S., Rhoads, J.~E., et al.\ 2020, \apj, 893, 134

\bibitem[Kroupa(2001)]{kro01} Kroupa, P.\ 2001, \mnras, 322, 231

\bibitem[Kunth \& {\"O}stlin(2000)]{kun00} Kunth, D., \& {\"O}stlin, G.\ 2000, \aapr, 10, 1 

\bibitem[Lian et al.(2016)]{lia16} Lian, J., Hu, N., Fang, G., Ye, C., \& Kong, X.\ 2016, \apj, 819, 73 

\bibitem[Maiolino et al.(2008)]{mai08} Maiolino, R., Nagao, T., Grazian, A., et al.\ 2008, \aap, 488, 463

\bibitem[Maiolino \& Mannucci(2019)]{mai19} Maiolino, R. \& Mannucci, F.\ 2019, \aapr, 27, 3

\bibitem[Mannucci et al.(2010)]{man10} Mannucci, F., Cresci, G., Maiolino, R., Marconi, A., \& Gnerucci, A.\ 2010, \mnras, 408, 2115

\bibitem[Marigo \etal(2008)]{marigo+08} Marigo, P., Girardi, L., Bressan, A., \etal\ 2008, \aap, 482, 883 

\bibitem[Martin et al.(2005)]{mar05} Martin, D.~C., Fanson, J., Schiminovich, D., et al.\ 2005, \apjl, 619, L1

\bibitem[Maseda et al.(2018)]{mas18b} Maseda, M.~V., Bacon, R., Franx, M., et al.\ 2018, \apjl, 865, L1 

\bibitem[McQuinn et al.(2020)]{mcq20} McQuinn, K.~B.~W., Berg, D.~A., Skillman, E.~D., et al.\ 2020, \apj, 891, 181. doi:10.3847/1538-4357/ab7447

\bibitem[Moustakas et al.(2006)]{mou06} Moustakas, J., Kennicutt, R.~C., \& Tremonti, C.~A.\ 2006, \apj, 642, 775. doi:10.1086/500964

\bibitem[Muzzin et al.(2013)]{muz13} Muzzin, A., Marchesini, D., Stefanon, M., et al.\ 2013, \apj, 777, 18

\bibitem[Nagao et al.(2006)]{nag06} Nagao, T., Maiolino, R., \& Marconi, A.\ 2006, \aap, 459, 85 

\bibitem[Noll et al.(2009)]{nol09} Noll, S., Burgarella, D., Giovannoli, E., et al.\ 2009, \aap, 507, 1793

\bibitem[Osterbrock(1989)]{ost89} Osterbrock, D.~E.\ 1989, Mill Valley, CA, University Science Books, 422 p., 

\bibitem[P{\'e}rez-Montero et al.(2013)]{per13} P{\'e}rez-Montero, E., Contini, T., Lamareille, F., et al.\ 2013, \aap, 549, A25

\bibitem[Pettini \& Pagel(2004)]{pet04} Pettini, M., \& Pagel, B.~E.~J.\ 2004, \mnras, 348, L59 

\bibitem[Pilyugin \& Thuan(2005)]{pil05} Pilyugin, L.~S., \& Thuan, T.~X.\ 2005, \apj, 631, 231 

\bibitem[Pirzkal et al.(2004)]{pir04} Pirzkal, N., Xu, C., Malhotra, S., et al.\ 2004, \apjs, 154, 501 

\bibitem[Salpeter(1955)]{sal55} Salpeter, E.~E.\ 1955, \apj, 121, 161

\bibitem[Salzer et al.(2000)]{sal00} Salzer, J.~J., Gronwall, C., Lipovetsky, V.~A., et al.\ 2000, \aj, 120, 80 

\bibitem[S{\'a}nchez et al.(2012)]{san12} S{\'a}nchez, S.~F., Kennicutt, R.~C., Gil de Paz, A., et al.\ 2012, \aap, 538, A8 

\bibitem[Sargent \& Searle(1970)]{sar70} Sargent, W.~L.~W., \& Searle, L.\ 1970, \apjl, 162, L155 

\bibitem[Schlafly \& Finkbeiner(2011)]{sch11} Schlafly, E.~F. \& Finkbeiner, D.~P.\ 2011, \apj, 737, 103. doi:10.1088/0004-637X/737/2/103

\bibitem[Tremonti et al.(2004)]{tre04} Tremonti, C.~A., Heckman, T.~M., Kauffmann, G., et al.\ 2004, \apj, 613, 898 

\bibitem[van Dokkum(2001)]{van01} van Dokkum, P.~G.\ 2001, \pasp, 113, 1420

\bibitem[van Dokkum et al.(2011)]{van11} van Dokkum, P.~G., Brammer, G., Fumagalli, M., et al.\ 2011, \apjl, 743, L15 

\bibitem[Wright et al.(2010)]{wri10} Wright, E.~L., Eisenhardt, P.~R.~M., Mainzer, A.~K., et al.\ 2010, \aj, 140, 1868

\bibitem[Yang et al.(2017a)]{yan17} Yang, H., Malhotra, S., Rhoads, J.~E., \& Wang, J.\ 2017a, \apj, 847, 38

\bibitem[York et al.(2000)]{yor00} York, D.~G., Adelman, J., Anderson, J.~E., et al.\ 2000, \aj, 120, 1579

\end{thebibliography}



\end{document}